\documentclass[reprint,superscriptaddress,showpacs,amssymb,floatfix]{revtex4-1}
\usepackage[final]{graphicx}
\usepackage{amsmath}
\usepackage{siunitx}
\usepackage[mathscr]{euscript}

\renewcommand{\d}{\mathrm{d}}

\begin{document} 

\title{Spin-orbit interaction of light in three-dimensional microcavities}

\author{Jakob Kreismann}
\email{jakob.kreismann@tu-ilmenau.de}
\affiliation{Institute for Physics, Theoretical Physics II/Computational Physics Group, Technische Universit\"{a}t Ilmenau, Weimarer Stra\ss{}e 25, 98693 Ilmenau, Germany}
\author{Martina Hentschel}
\affiliation{Institute for Physics, Faculty of Natural Sciences, Technische Universit\"{a}t Chemnitz, Reichenhainer Str. 70, 09126 Chemnitz, Germany}

\date{\today}

\begin{abstract}
\textbf{We investigate the spin-orbit coupling of light in three-dimensional cylindrical and tube-like whispering gallery mode resonators. We show that its origin is the transverse confinement of light in the resonator walls, even in the absence of  inhomogeneities or anisotropies. 
The spin-orbit interaction results in elliptical far-field polarization (spin) states and causes spatial separation of polarization handedness in the far field. The ellipticity and spatial separation are enhanced for whispering gallery modes with higher excitation numbers along the resonator height. We analyze the asymmetry of the ellipticity and the tilt of the polarization orientation in the far field of cone-like microcavities.   
Furthermore, we find a direct relationship between the tilt of the polarization orientation in the far field and the local inclination of the resonator wall.
Our findings are based on FDTD-simulations and are supported by three-dimensional diffraction theory.}
\end{abstract}


\maketitle

 Mircooptical systems which confine light to small volumes have received a lot of interest in the past years~\cite{Chang1996,Vahala2003}. Well known examples of such systems are e.g.  microspheres~\cite{Collot1993,Gorodetsky2000,VollmerYang2012}, microtoroids~\cite{Ilchenko2001,Armani2003} and microdisks~\cite{McCall1992,Michael2007}. In particular, bottle-like and tube-like microcavities have received much attention in recent years ~\cite{wgm_bottles2004,wgm-rings_2007,wgm-bottles_2008,wgm-bottles_2009,wgm-bottles_2012,spinorbit-cones_2016,wgm-tubes_2017,wgm-tubes_2019}. In contrast to rather flat microdisks, these types of cavities allow a full three-dimensional (3D) formation of the resonances. 3D whispering gallery modes (WGMs) have been theoretically studied in e.g.~\cite{wgm_bottles2004,Lacey_2001,Schwefel_2005,Teraoka_2006,Kreismann2017,Kreismann2018,Gladyshev_2018,KhosraviSO_2019}.\\
One aspect of interest is the polarization evolution in 3D microcavities, where, unlike the two-dimensional situation, the polarization directions do not decouple. This enables a coupling between the light's orbital motion in the resonator and its polarization (spin of light) state that is known as spin-orbit interaction of light~\cite{BliokhSO_2015,AielloS0_2015,CardanoSO_2015}. It has been studied in different contexts~\cite{JungeSO_2013,OConnorSO_2014,ShaoSO_2018,spinorbit-beams_2019,RosenbergerSO_2019,KhosraviSO_2019,SunSO_2019}, and a special focus was given to asymmetric microcavities and the role of anisotropies or inhomegeneties~\cite{spinorbit-cones_2016} as well as the interpretation in terms of geometric phases~\cite{Berry1984,Berry1987,Bliokh2008,BliokhGP_2019}. 

Here, we investigate spin-orbit interaction of light in symmetric and asymmetric photonic microsystems that are deduced from ring-like (hollow-cylinder type) microcavities. The generic resonances are known to be whispering-gallery type modes. The focus of this manuscript is the investigation of their spin-orbit interaction in dependence on the resonator geometry: How is the polarization state of light affected by inclining the resonator wall and manipulating its thickness, and what role plays the resonance morphology/excitation number ? Special attention will be given to the far-field polarization properties as this allows for a direct observation of our findings and their use in potential applications such a sensors or polarizers.\\ 
Whereas  numerical finite difference time domain (FDTD) simulations will play a major role throughout the paper, we shall see that optics in form of Kirchhoff's diffraction theory yields valuable insight and understanding of the simulation results. This implies, however, that an explanation based geometric phases cannot be the objective of this paper. Though we will see manifold examples of the interplay between the resonator geometry and the resulting polarization evolution of light throughout this paper, the well-known explanation in terms of geometric phases and solid angles spanned in parameter space (that applies also e.g. to spin-dependent transport of electrons along rings subject to inhomogeneous magnetic fields~\cite{Loss_1990,Nagasawa_2013,Frustaglia2001,Hentschel2004}) is not sufficient to capture the more complex situation that includes transformation into the far field we are interested in here.\\  
The paper is organized as follows. In the first section, we will recap the theory of spin-orbit interaction of light applied to a whispering gallery mode (WGM) in a 3D dielectric ring resonator and, in section 2, apply it 
to an azimuthally propagating mode and its far-field emission. In the third section, we will study 3D WGMs in cone-like tube cavities.
As in the previous section, we will investigate the far-field polarization states and explain differences to the previous case. In the following section IV, we study the role of inhomogeneous resonator wall structures and finish the paper with a summary. At the end of this paper, we give a short description of the used FDTD-method. The results of vector diffraction theory that are used throughout this work are explained and summarized in a Supplemental Material (SM).

\section{Spin-orbit coupling of light} 
\label{sec:theory}

We start the theoretical description of 3D optical microcavities by recapping  Gauss\textquoteright s law in differential form~\cite{Jackson} for time-harmonic fields in dielectric media without free electric charges $\nabla\cdot\left(\varepsilon(\textbf{r})\textbf{E}(\textbf{r}) \right)=0$, where $\varepsilon(\textbf{r})$ describes an inhomogeneous dielectric material permittivity, $\textbf{E}(\textbf{r})$ represents the electric field vector and $\textbf{r}$ is the spatial coordinate. Applying the chain-rule and rearranging terms leads to
\begin{equation}
	\nabla\cdot\textbf{E}= -\frac{1}{\varepsilon(\textbf{r})}\left(\nabla\varepsilon(\textbf{r}) \right)\cdot \textbf{E}.
	\label{eq1}
\end{equation}

This implies that the divergence of the electric field does not generally vanish (as it does in vacuum or homogeneous materials). Rather, it will take a finite value that depends on the orientation of the electric field with respect to the gradient of the permittivity. In cylindrical  coordinates, Eq.(\ref{eq1}) reads
\begin{equation}
	\frac{1}{r}\partial_r (r E_\text{r}) + \frac{1}{r}\partial_\phi E_\phi + \partial_z E_\text{z} = -\frac{1}{\varepsilon(\textbf{r})}\left(\varepsilon'_\text{r} E_\text{r} + \frac{\varepsilon'_\phi}{r} E_\phi +\varepsilon'_z E_\text{z}  \right) \nonumber
\end{equation}
with $r$ the distance from the (cylinder) $z$-axis, $\partial_r$ the partial derivative with respect to $r$ and $\varepsilon'_{r(\phi,z)}=\partial_{r(\phi,z)}\varepsilon(\textbf{r})$. 

\begin{figure}
	\centering
	\includegraphics[width=8.5cm]{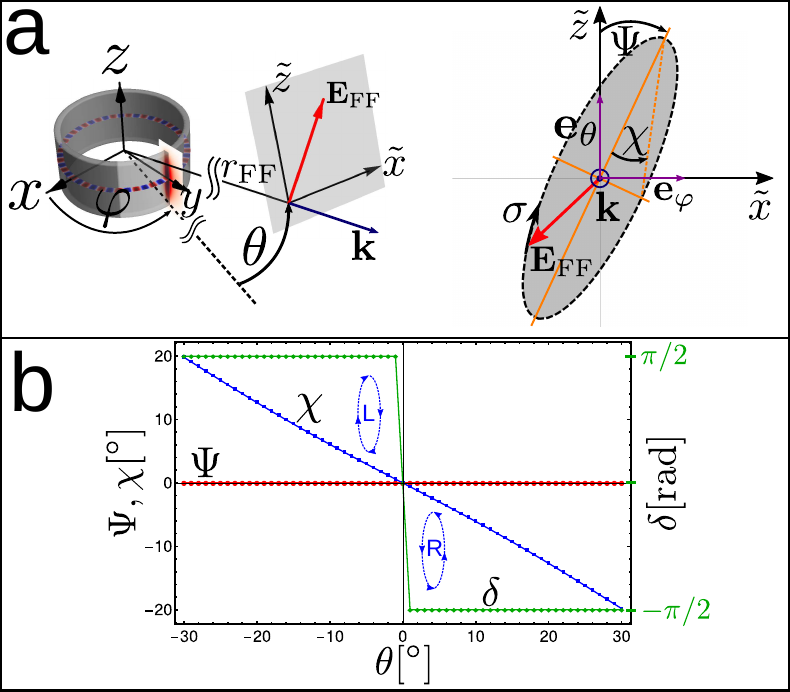}
	\caption{\textbf{Far-field polarization properties.} \textbf{a:} Characterization of the far-field polarization state by $\Psi$ - angle of orientation, $\chi$ - ellipticity angle and $\sigma$ - handedness. \textbf{b:} $\Psi$ and $\chi$ of a 3D-ring WGM far field derived from the Kirchhoff diffraction formula for the fundamental axial mode ($q=1$). See text for details. }
	\label{fig:f0a}
\end{figure}

We now investigate a mode that is propagating into $\phi$-direction, cyclically guided by a thin dielectric ring with constant mean radius $r=R$. Hence, the $\phi$-component is the longitudinal component, whereas the $r$- and $z$-components correspond to transverse (tr) components. Sorting by components yields:
\begin{equation}
	\frac{1}{R}\left(\partial_\phi E_\phi + \frac{\varepsilon'_\phi}{\varepsilon} E_\phi\right) = -\left(\nabla_\text{tr}\cdot \textbf{E}_\text{tr} + \frac{1}{\varepsilon} \left(\nabla_\text{tr}\varepsilon\right)\cdot \textbf{E}_\text{tr} \right).
	\label{equ:longi_comp_1}
\end{equation}
%
%
We see that the longitudinal component $E_\phi$ and its change depend on the transverse confinement (first term on the right-hand side) and the transverse gradient of the material permittivity (second term on the right-hand side). In other words, an initially purely transverse field can induce a longitudinal component. This action of the light field orbit on its overall polarization is known as spin-orbit interaction. \\
The complete electric field vector reads 
\begin{align}
\textbf{E}(r,\phi,z) &= \left(E_\text{r}\textbf{e}_\text{r} + E_\phi\textbf{e}_\phi + E_\text{z}\textbf{e}_\text{z} \right) \\
 &= \left(A_\text{r}(r,z)\textbf{e}_\text{r} + A_\phi(r,z)\textbf{e}_\phi + A_\text{z}(r,z)\textbf{e}_\text{z} \right) e^{i m \phi}
\end{align}
where the amplitudes $A_\text{r}$ and $A_\text{z}$ represent the transverse mode profile, $A_\phi$ is the longitudinal amplitude and $m$ is the azimuthal mode number. The $e^{i m \phi}$-factor indicates a azimuthally traveling wave.\\
Applying Eq.~(\ref{equ:longi_comp_1}) to the complete electric field vector of the mode and expanding a fraction yields
\begin{equation}
E_\phi=i \frac{R}{m} \frac{1}{1 - i \xi}
\left(\nabla_\text{tr}\cdot \textbf{E}_\text{tr} + \frac{1}{\varepsilon} \left(\nabla_\text{tr}\varepsilon\right)\cdot \textbf{E}_\text{tr} \right)
\label{equ:Ez_beam}
\end{equation}
where $\xi=\frac{1}{m}\frac{\varepsilon'_\phi}{\varepsilon}$ 
is a parameter describing how strong the material properties (components of the permittivity $\epsilon$ or the related refractive index $n$ with $\varepsilon=n^2$) change along the $\phi$-direction with respect to the azimuthal mode number $m$.
\noindent Eq.~(\ref{equ:Ez_beam}) describes the spin-orbit interaction of light in WGMs.
The orbital momentum of the mode represented by $m$ is transformed into a (phase-shifted) spin-momentum -- the polarization -- represented by $E_\phi$.  
The spin-orbit interaction depends on the transverse confinement $\nabla_\text{tr}\cdot\textbf{E}_\text{tr}$, the transverse material  gradient $\nabla_\text{tr}\varepsilon$, and material change $\xi$ along the propagation direction $\phi$. 
In particular, starting from a TE-like WGM where the transverse electric field is  aligned (almost) parallel to the resonator wall, $\textbf{E}_\text{tr}\approx E_\text{z}\textbf{e}_\text{z} $, we find that spin-orbit coupling induces an  $E_\phi$ component proportional to the derivative of $E_\text{z}$, $E_\phi\propto\partial E_{\text{z}} /\partial z $.\\
If there is no ($\xi=0$) or only weak material change ($\xi\ll 1 $) along the $\phi$-direction, the prefactor reduces to $i R/m$. That is, the longitudinal component undergoes a phase shift of $\pi/2$ or a factor $i$. As a consequence, the propagating mode is elliptically polarized with the polarization ellipse lying in a plane spanned by $\textbf{e}_\text{z}$ and $\textbf{e}_\phi$.\\
We will now briefly analyze  under which conditions an originally linearly polarized mode can reach the limiting case of circular polarization, that is, equal longitudinal and transverse amplitudes $\left|E_\phi\right|\sim\left|\textbf{E}_\text{tr}\right|$. This requires 
\begin{equation}
	\frac{1}{2\pi}\frac{\lambda}{n}
	\frac{\nabla_\text{tr}\cdot\textbf{E}_\text{tr}}{\left|\textbf{E}_\text{tr}\right|}\overset{!}{=}1 
	\quad\text{or}\quad 
	\frac{1}{2\pi}\frac{\lambda}{n}
	\frac{\left|\nabla_\text{tr}\varepsilon\right|}{\varepsilon}\overset{!}{=}1
	\label{equ:circ-pol_cond}
\end{equation} 
where we approximated the term $R/m$, cf.~Eq.~[\ref{equ:Ez_beam}], by $\lambda/2\pi n$ by applying the resonance condition $2\pi R \approx m\lambda /n$ of a WGM with $\lambda/n$ being the medium wavelength. In order to fulfill the first condition of Eq.~(\ref{equ:circ-pol_cond}), the normalized change of the amplitude of the transverse electric field on the medium wavelength scale has to be of the order of $2\pi$. This corresponds, however, to confinement of light on a scale of the medium wavelength. Such a strong confinement 
can be realized by highly focused beams as recently investigated in~\cite{spinorbit-beams_2019}, or WGM resonances in tube-like or bottle-like cavities, see~\cite{wgm-bottles_2008,wgm-bottles_2009,wgm-bottles_2012,wgm-rings_2007,wgm-tubes_2009,Belanos_2012,wgm-tubes_2017,wgm-tubes_2019,wgm_bottles2004}.\\
The second condition of Eq.~(\ref{equ:circ-pol_cond}) requires the normalized transverse permittivity of the material to change by a factor of $2\pi\varepsilon$ over one medium wavelength. Even for low refractive indices like $n=1.2$ ($\varepsilon=1.44$), this requires a permittivity change by the factor of $9$ on the wavelength scale. At interfaces of different material, strong material gradients may occur but these gradients are localized in the region of the material interface. For pure dielectric materials, a continuously strong permittivity change requires strong material inhomogeneities.
We leave this to another study and focus here on homogeneous materials with strong transverse confinement of light such as in the walls of 3D ring resonators. We shall see that already this generic situation can induce substantial spin-orbit coupling with a number of different effects.\\ 
This applies in particular to the observation of light in the far field. Therefore, we link now the fields inside the resonator to the far-field polarization states.  Note that the electromagnetic radiation $\textbf{E}_\text{FF}$ at distance $r_\text{FF}$ in the far field 
has to be an outgoing spherical wave,
\begin{equation}
\textbf{E}_\text{FF}=\left(E_\varphi \textbf{e}_\varphi + E_\theta \textbf{e}_\theta  \right)\exp{(i\textbf{k}\cdot\textbf{r}_\text{FF} - i \omega t)},
\end{equation}
where $\textbf{e}_\varphi$ and $\textbf{e}_\theta$ are the unit vectors in direction of $\varphi$ and $\theta$, respectively, as known from spherical coordinates. Both are perpendicular (transverse) to the propagation direction $ \textbf{k}/|\textbf{k}|$. $E_\varphi$ and $E_\theta$ are the corresponding $\varphi$- and $\theta$-components. \\
In a local far-field coordinate system $(\tilde{x},\tilde{z})$, where the $\tilde{x}$-axis and $\tilde{z}$-axis are spanned by  $\textbf{e}_\varphi$ and $\textbf{e}_\theta$, respectively, the physically observable electric field given by the real part of $\textbf{E}_\text{FF}$ describes an ellipse in general, see Fig.~\ref{fig:f0a} \textbf{a}. This polarization ellipse is characterized by (i) the orientation angle $\Psi$ of its major axis with respect to the $\tilde{z}$-axis, (ii) the ellipticity angle $\chi$ (with $\tan  \chi =$ minor axis/major axis)  and (iii) the handedness $\sigma$ of the direction of rotation of $\text{Re}[\textbf{E}_\text{FF}]$  when looking against the propagation direction (looking towards the resonator), see Fig.~\ref{fig:f0a}.\\
The quantities $\Psi$, $\chi$ and $\sigma$ are given by~\cite{polarization_ellipse}:
\begin{align}
	\tan 2\Psi &= \frac{2\nu}{1-\nu^2}\cos\delta \label{equ:def_Psi}   \\
	\sin 2\chi &= \frac{2\nu}{1+\nu^2}\sin\delta \label{equ:def_chi}   \\
	\sigma &= \text{sign}(\delta) \label{equ:def_sigma}
\end{align}
where $\nu=A_\varphi/A_\theta$ is the ratio of the amplitudes of the far-field components $E_\varphi$ and $E_\theta$, and $\delta=\text{Arg}(E_\varphi/E_\theta)$ represents the phase difference between $E_\varphi$ and $E_\theta$. 
The angle $\Psi$ describes the tilt with respect to the local $\tilde{z}$-axis and ranges from $-90^\circ$ to $+90^\circ$.  $\Psi=0^\circ$ and $\Psi=90^\circ$ correspond to vertical (parallel to the $\tilde{z}$-axis) and horizontal (parallel to the $\tilde{x}$-axis) polarization orientation, respectively.
The ellipticity angle $\chi$ describes the states of linear ($\chi=0^\circ$), elliptical ($0^\circ < |\chi| < 45^\circ$) and circular ($|\chi|=45^\circ$) polarization. The sign of $\chi$ corresponds to the sign of the phase difference $\delta$ because of the $\sin\delta$-term in Eq.~(\ref{equ:def_chi}). $\chi<0$ and $\chi>0$ represent right-handed ($\sigma=-1$)  and left-handed ($\sigma=+1$) polarization, respectively.\\
\noindent The desired relation between the fields inside the ring resonator $(E_\phi,E_\text{z})$ and the far-field components $(E_\varphi,E_\theta)$ is provided by the Kirchhoff diffraction theory that we explain in more detail in the Supplemental material. To this end we treat the side wall of the 3D ring resonator as aperture that diffracts the fields of the WGM resonances into the far field. Due to the thin-wall structure, we approximate  
the 3D-Ring by a cylinder surface $S$ of radius $R$ and height $h$. The electric far-field vector $\textbf{E}_\text{FF}$ can be computed by evaluating Kirchhoff's vector diffraction formula in the Fraunhofer (or far-field) limit ~\cite{Jackson} yielding
\begin{equation}
	\textbf{E}_\text{FF} \sim \textbf{k}\times\iint_S \textbf{n}'\times\textbf{E}(\textbf{x}')\exp{(-i \textbf{k}\cdot \textbf{x}')}\text{d}a' \:,
	\label{eq_kirchhoff_diff}
\end{equation}
where $\textbf{x}'$ is the position vector on the diffracting surface $S$ with area element $\text{d}a' = \text{d} z' \, R \text{d}\phi'= h \text{d}u' \, R\text{d}\phi'$ and electric field $\textbf{E}(\textbf{x}')$, $\textbf{n}'$ is the unit vector normal to the surface, and $\textbf{k}$ the far-field wave vector. Note that only fields parallel to the surface $S$ (represented by the term $\textbf{n}'\times\textbf{E}(\textbf{x}')$ in Eq.~(\ref{eq_kirchhoff_diff}) are considered. Contributions from normal field components, $\propto \textbf{n}' \textbf{E}(\textbf{x}')$, which would show up as an additional magnetic field term inside Eq.~(\ref{eq_kirchhoff_diff}), 
were neglected because the predominantly TE-like character of the WGM resonances ensures the electric field components to be parallel to the resonator wall and thus parallel to the diffracting surface $S$. 
The fields at the cylindrical diffracting surface can be represented as $\textbf{E}(\textbf{x}') = \left(A_\text{z}(z')\textbf{e}_\text{z} + A_\phi(z')\textbf{e}_\phi \right)\exp{(im\phi')}$.\\
Evaluating the diffraction formula Eq.~(\ref{eq_kirchhoff_diff}), we obtain the sought-after expressions for the far-field components $E_\varphi$ and $E_\theta$ (see SM),
\begin{align}
E_\varphi(\varphi,\theta)		={} & h R \exp{(im\varphi)}\exp{\left(-i\frac{\pi}{2}m\right)} J_m\left(\tilde{k}_1\right)  \nonumber \\
             &{} \times\bigg(\hat{A}_\text{z}(\tilde{k}_2)\frac{\tilde{k}_1^2-m^2}{m\tilde{k}_1}\sin\theta - \Delta\cos\theta \bigg) \label{equ:EphiFF} \\[8pt]
E_\theta (\varphi,\theta)		={} & h R \exp{\left(im\varphi\right)}\exp{\left(-i\frac{\pi}{2}(m+1)\right)} \nonumber\\
			&{}
			\times\hat{A}_\text{z}(\tilde{k}_2) \frac{J_{m-1}\left(\tilde{k}_1\right)-J_{m+1}\left(\tilde{k}_1\right)}{2} \label{equ:EtheFF}
\end{align}
where the $J_\mu$ are the Bessel functions of the first kind with $\tilde{k}_1=kR\cos\theta$ and $\tilde{k}_2=kh\sin\theta$. $\hat{A}_\text{z}(\tilde{k}_2)$ and $\Delta$ represent the following expressions using $\d z'=h\d u'$:
\begin{align}
\hat{A}_\text{z}(\tilde{k}_2) ={} & \int_{-1/2}^{1/2}\tilde{A}_\text{z}(u') \exp{(-i\tilde{k}_2 u')} \text{d}u' \label{equ:AzFF}  \\[5pt]
\Delta  ={} &  \frac{i}{m}\frac{R}{h}\tilde{A}_\text{z}(u')\exp{\left(-i\tilde{k}_2 u'\right)}\bigg|_{u_1'=-1/2}^{u_2'=1/2} \nonumber  \\[5pt] 
		={} & \frac{2R}{m h}\tilde{A}_\text{z}(1/2) \cdot \begin{cases}
		\sin\left(\tilde{k}_2/2\right) &\text{for $q=1,3,5,...$} \label{equ:Delta}  \\[10pt]
		i\cos\left(\tilde{k}_2/2\right) &\text{for $q=2,4,6,...$}
		\end{cases}
\end{align}
where Eq.~(\ref{equ:AzFF}) resembles the diffraction pattern of a slit of height $h$ with the local amplitude $\tilde{A}_\text{z}(u')$. Due to the strong confinement, the amplitude $\tilde{A}_\text{z}(u')$ rapidly converges towards zero above and below the wall of the ring as shown in Fig.~\ref{fig:f1} and, therefore, $\hat{A}_\text{z}(\tilde{k}_2)$ can be interpreted as the Fourier transform of $\tilde{A}_\text{z}(u')$. 
Eq.~(\ref{equ:Delta}) describes an additional term taking the amplitude $\tilde{A}_\text{z}$ at the upper and lower boundary into account where $q$ represents the axial mode number, cf sec.~\ref{sec:3Dring}. It arises from integration by parts of the $\tilde{A}_\phi(u')$ component ( $\tilde{A}_\phi\propto \partial \tilde{A}_\text{z}/\partial u'$), cf. supplemental material.\\  
For the sake of simplicity, we neglect the boundary amplitude in Eq.~(\ref{equ:Delta}) for the fundamental axial mode ($q=1$), $\tilde{A}_z(1/2)=\tilde{A}_z(-1/2)\approx0$. As a result, $\Delta=0$ and
we see that the $E_\varphi$ and $E_\theta$ components of the far field (Eqs.~\ref{equ:EphiFF} and \ref{equ:EtheFF}) are directly connected to the Fourier transform of the $E_\text{z}$ component of the WGM field at the resonator wall. Furthermore, the $E_\varphi$ component changes its sign at $\theta=0$ because of the $\sin\theta$ term which indicates far-field polarization states of opposite handedness for $\theta<0$ and $\theta>0$.
Remarkably, the polarization quantities $\Psi$ and $\chi$ of the fundamental axial mode ($q=1$) are independent of the far-field pattern represented by $\hat{A}_\text{z}(\tilde{k}_2)$ because this term cancels out in the parameter $\nu$ and it is also irrelevant for the phase difference $\delta$. Thus, the spatial splitting of the handedness of the far-field polarization states, as illustrated in Fig.~\ref{fig:f0a} \textbf{b}, is an intrinsic feature of a propagating (non-standing) WGM.\\
For higher axial modes ($q>1$), the contribution of $\Delta$ increases because of increasing boundary amplitudes, cf. Fig.~\ref{fig:f1}, and will enhance the ellipticity $\chi$ as shown in the results further below. The fundamental feature of spatial splitting of the far-field polarization handedness is also enhanced.

\section{Cylindrical 3D-ring resonators}
\label{sec:3Dring}
\begin{figure}
	\centering
	\includegraphics[width=8.5cm]{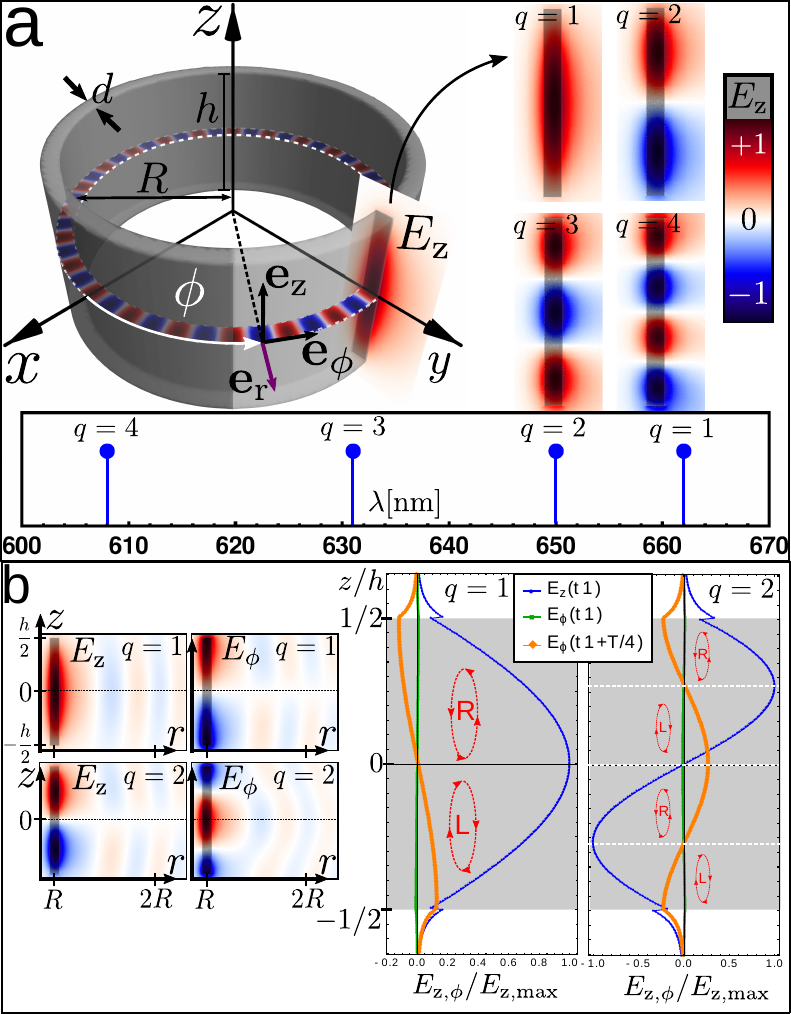}
	\caption{\textbf{Geometry and WGM.} \textbf{a:} Illustration of a cylindrical ring of thickness $d$ and corresponding WGMs, and spectrum of the modes with different vertical excitations. Parameters: azimuthal number  $m=24$, mean radius $R=2\mu m$, height $h= R = 2\mu m$, wall thickness $d=0.2\mu m$ and refractive index $n=1.5$. \textbf{b:} (left-hand side) field distribution of $E_{\text{z}}$ and $E_\phi$ in the exterior space ($r>R$) of the ring. (right-hand side)  $E_{\text{z}}$ and $E_\phi$ at $r=R$ as a function of $z$. See text for details. }
	\label{fig:f1}
\end{figure}

We begin our study 
of spin-orbit interaction of light in a 3D-ring resonator cavity with mean radius $R$, wall thickness $d$ and height $h$. The electromagnetic eigenmodes of such cavities are WGMs (hosted in the cavity cross section) subject to the additional confinement in $z$-direction, see Fig.~\ref{fig:f1} a. The field pattern of $E_\text{z}$ exhibits different types of excitations that can be characterized by the axial mode number $q$ representing the number of extrema  of $E_\text{z}$ along the resonator height.\\
\noindent Fig.~\ref{fig:f1} \textbf{a} illustrates the ring geometry and slices of the 3D field distribution $E_\text{z}$ at $z=0$ (showing the radial distribution of the field) and at $x=0$ (showing its vertical distribution).  
The field distributions are taken from a temporal snapshot of a clock-wise propagating wave with $m=24$ and $q=1$. The panel on the right-hand side shows the fundamental vertical excitation $q=1,2,3,4$ and the lower panel 
their spectral positions, indicating that the wavelength decreases with increasing excitation as expected.\\
The diagrams on the right-hand side of Fig.~\ref{fig:f1} \textbf{b} show the transverse $E_{\text{z}}$ and the longitudinal
component $E_\phi$ of the electric field of the fundamental mode $q=1$ and the first vertical excitation $q=2$ at two different time steps, $t_1$ and $t_1 + \frac{T}{4}$ with $T$ being the optical period. The fields are taken along the center of the wall ($r=R$), c.f. the insets on the left-hand side.\\
We see that the graphs of $E_\phi$ correspond to the derivative of $E_\text{z}$, cf.~Eq.~(\ref{equ:Ez_beam}). Furthermore, we see that $E_\phi$ reaches its maximum a quarter period later than $E_{\text{z}}$ because of the factor $i$ resulting from the confinement of the propagating mode, cf.~Sec.~1.  This phase shift of $\pi/2$  generates elliptical polarization inside the ring resonator as illustrated by the red ellipses. Note that the polarization ellipse lies in a plane spanned by $\textbf{e}_\text{z}$ and $\textbf{e}_\phi$, that is the polarization ellipse lies parallel to the propagation direction.
The upper (lower) half of the $q=1$-WGM carries right (left)-handed elliptical polarization where right (left)-handed elliptical or circular polarization is defined by opposite (same) sign of the $E_\text{z}$ and $E_\phi$ components.\\ 
As we shall see below, precisely this splitting is transferred into the far-field and can be observed there. The $q=2$-WGM has 4 regions (separated by horizontal dashed lines in Fig. \ref{fig:f1} b) of alternating right and left-handed polarization. We point out that elliptical polarization 
occurs in traveling waves only whereas in standing wave WGMs the transverse spin-momenta of the counter-propagating modes cancel exactly and yield linear polarization.\\
In the next step, we will investigate how the far-field polarization state depends i) on the far-field (observation) angle and ii) on the vertical excitation number $q$.
\begin{figure}
	\centering
	\includegraphics[width=8.5cm]{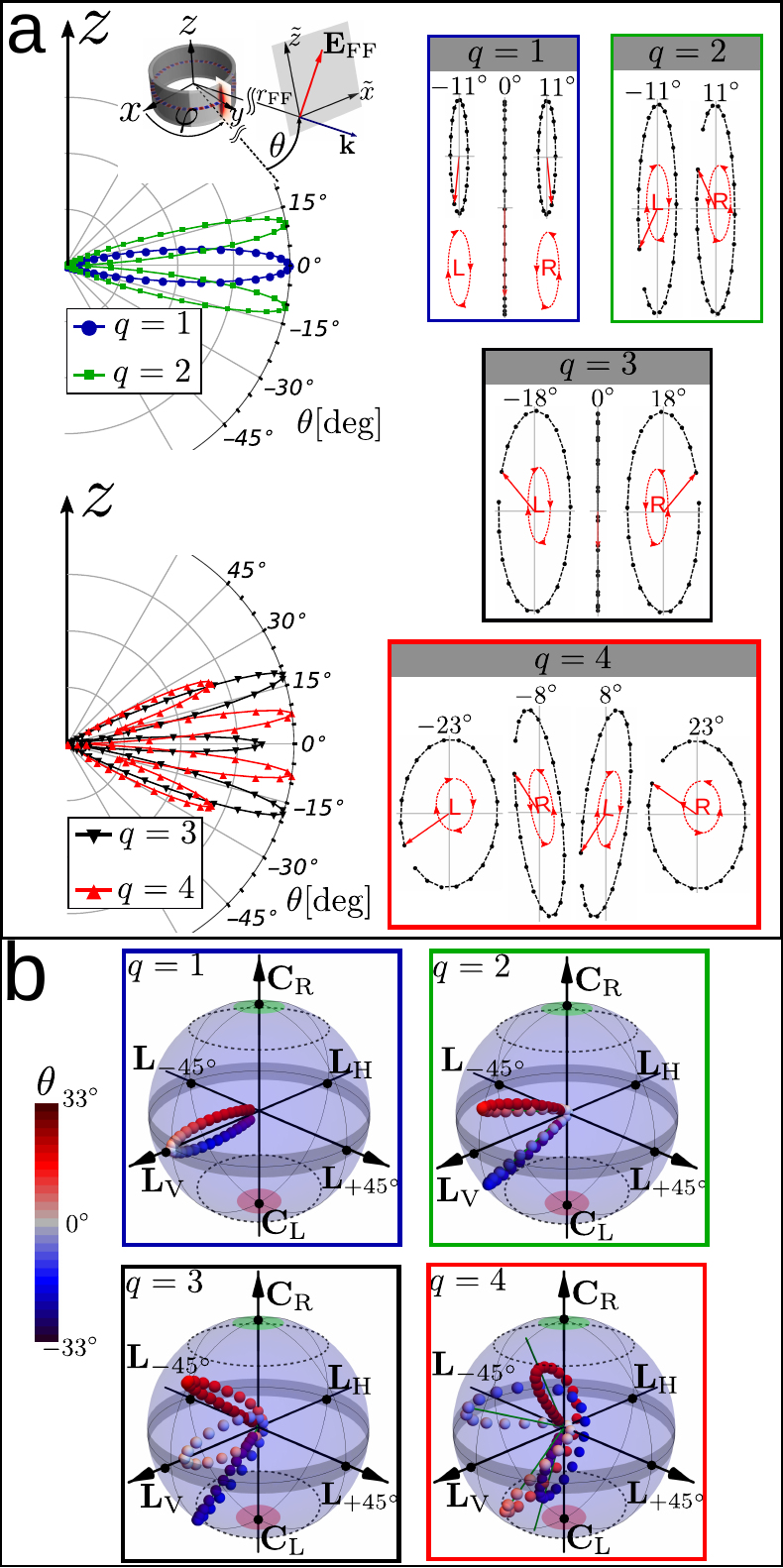}
	\caption{ \textbf{Far fields and far-field polarization states.} \textbf{a:} Polar plots of far-field intensities of different vertical excitation numbers $q$. The frames show the polarization ellipses along the (maximum) far-field direction $\theta$ indicated above each ellipse. 
 \textbf{b:} Poincar\'{e}-spheres of polarization displaying observable polarization states when scanning the far field within the given $\theta$-range.  See text for details.  }
	\label{fig:f2}
\end{figure}
The far fields were obtained by computing the far-field electric field vectors $\textbf{E}^{\text{FF}}$ and its far-field components $E_\varphi$ and $E_\theta$ in a distance of $r_\text{FF}=50\mu m$ from the origin, see inset of Fig.~\ref{fig:f2} \textbf{a}. The angles $(\varphi,\theta)$ are the azimuthal and elevation angle of the far field, respectively. The far-field polarization states are characterized by the orientation angle $\Psi$, the ellipticity angle $\chi$ and the handedness $\sigma$, cf. Fig.~\ref{fig:f0a} \textbf{a}.\\
Due to the rotational symmetry around the z-axis, the $\varphi$-dependence is trivial and thus we focus on the $\theta$-dependence of the far field. 
Fig.~\ref{fig:f2} \textbf{a} shows the far-field intensity $ \left| \textbf{E}^{\text{FF}}(\theta) \right|^2 $ as polar plot for $q=1,2$ and $q=3,4$, respectively. We find that $q$ corresponds directly to the number of observed far-field lobes. This connection can be explained by Fraunhofer (far-field) diffraction where the wall of the resonator acts as an aperture, see Eq.~(\ref{eq_kirchhoff_diff}). The dominant $E_\text{z}$-component of the WGM resonance will also rule the far-field pattern.\\
In Fig.~\ref{fig:f2} \textbf{a} the far-field electric field vector in the local $\tilde{x}$-$\tilde{z}$ coordinate system is shown at the elevation angles $\theta$ with maximum far field emission (and $\theta=\pm11^\circ$ for $q=1$) indicated on top of each frame. From top to bottom, the $q=1,2,3,4$ cases are shown. The electric field vectors are taken over one oscillation period where the red arrows and black dots correspond to the phase-dependent position of the first snapshot and the 19 following snapshots of the oscillation period, respectively.\\ 
In the case of the $q=1$-WGM, we observe left (right)-handed elliptical  polarization for negative (positive) far-field angles $\theta$ and linear polarization at the lobe maximum at $\theta=0$. This separation of the polarization handedness in the far-field arises from the reversed 
polarization distributions inside the resonator, cf. Fig.~\ref{fig:f1} \textbf{b}. The maximum of the lobe at $\theta=0^\circ$ shows linear polarization because the $z=0$-plane is a symmetry (mirror) plane at which the opposite handed polarization states compensate each other, resulting in linear polarization pointing into the $z$-direction, i.e. $\Psi = 0^\circ $. Similar results of polarization separation in the far field were observed for scattering of surface plasmons at nanostructures~\cite{OConnorSO_2014}.\\
The far field of the $q=2$-WGM has two pronounced lobes with opposite polarization handedness and stronger elliptical polarization compared to the $q=1$-case. The more pronounced elliptical polarization arises from the stronger confinement along the $z$-direction of the $q=2$ mode, and hence a stronger spin-orbit interaction generating a stronger $E_\phi$ component as shown in Fig.~\ref{fig:f1} \textbf{b} (maximum of $E_\phi/E_\text{z,max}$ for $q=2$ reaches almost $0.3$, where as $E_\phi/E_\text{z,max}$ for $q=1$ remains below $0.2$). The two far-field lobes arise from the dominant $E_\text{z}$ component, cf. Fig.~\ref{fig:f1} \textbf{b}, whereas the opposite handedness results from the opposite polarization distribution inside the resonator.\\
For $q=3$, the far field is very similar to the one of $q=1$ because in both cases it has an even $E_\text{z}$ and odd $E_\phi$ distribution inside the resonator. The stronger pronounced elliptical polarization arises from the now even stronger confinement as explained for $q=2$-case.\\
Eventually, the far field of the $q=4$-WGM shows 4 lobes of alternating polarization handedness and switching ellipticity. The outer lobes show almost circular polarization. The polarization orientation of the inner lobes is slightly tilted from the $z$-axis.\\
In order to further characterize the polarization states, we use their  
$\Psi$- and $\chi$-values to present them on the so-called Poincar\'{e}-sphere of polarization~\cite{polarization_ellipse}, cf.~Fig.~\ref{fig:f2} \textbf{b}. The sphere is spanned by $2\Psi$ and $2\chi$ which represent the azimuthal and elevation angle, respectively. The equator of this sphere ($\chi=0$) corresponds to linear polarization states of different orientations: $2\Psi=0$ - linear vertical ($\textbf{L}_\text{V}$), $2\Psi=\pi$ - linear horizontal ($\textbf{L}_\text{H}$), and $2\Psi=\pm \pi/2$ - linear inclined by $\pm 45^\circ$ ($\textbf{L}_{\pm45^\circ}$). The poles of the Poincar\'{e}-sphere at $2\chi=\pm \pi/2$ correspond to right-handed ($\textbf{C}_\text{R}$) and left-handed ($\textbf{C}_\text{L}$) circular polarization. All other states indicate elliptical polarization. The distance from the origin indicates the 
far-field intensity\\
The Poincar\'{e}-spheres in Fig.~\ref{fig:f2} \textbf{b} illustrate the observable far-field polarization states when scanning the far-field through the color-encoded $\theta$-range. For $q=1,2,3$ the polarization handedness is directly related to different spatial regions $\theta$ as is clearly visible by the blue (red) points remaining in the lower (upper) hemisphere.\\
We emphasize that the far-field polarization ellipse lies in a plane spanned by $\textbf{e}_\varphi$ and $\textbf{e}_\theta$ and is transverse to the propagation direction $\textbf{k}/|\textbf{k}|$, whereas the polarization ellipse inside the resonator is longitudinal to the propagation direction $\textbf{e}_\phi$. Thus, we observe a transition from a longitudinal to transverse polarization ellipse orientation in the far field.\\
\section{Cone-like 3D ring resonators}     
   
\begin{figure}
	\centering
	\includegraphics[width=8.5cm]{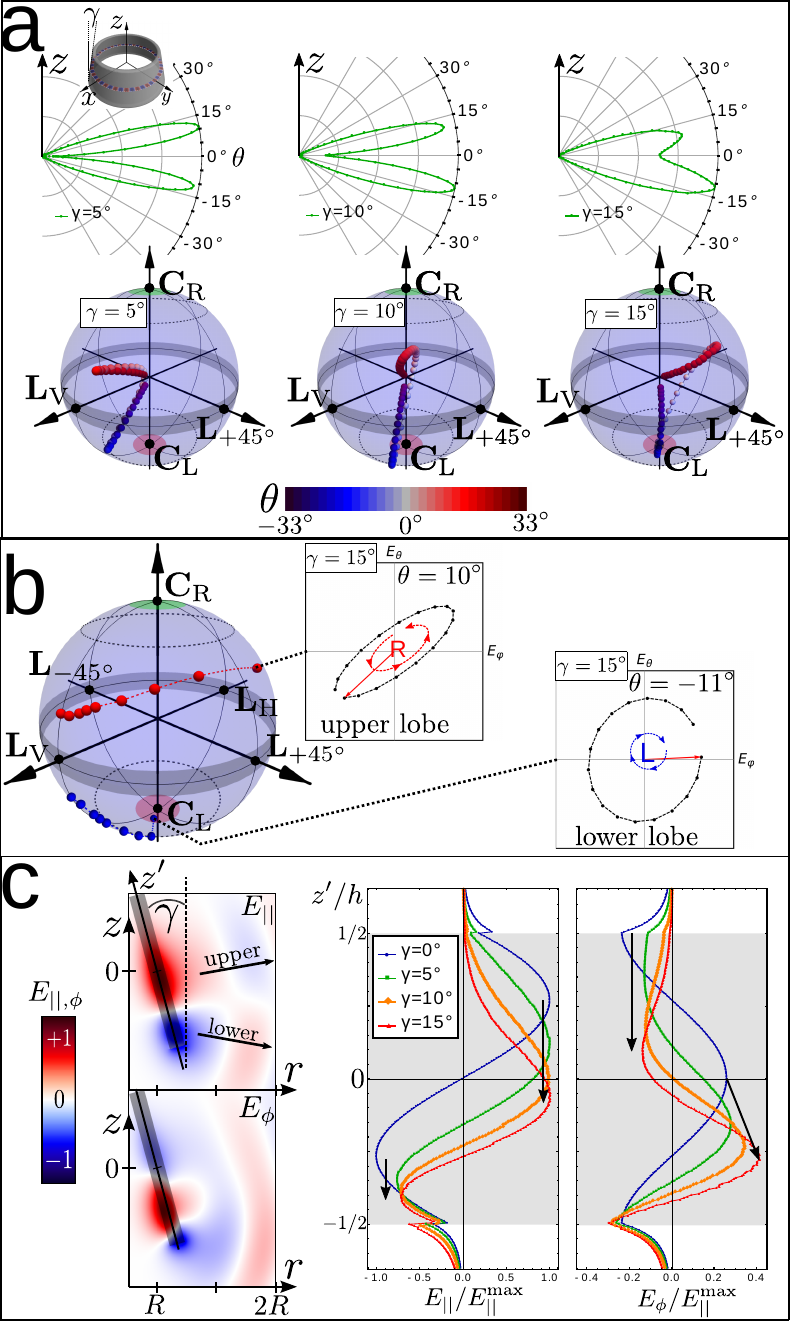}
	\caption{\textbf{Polarization states in cone-like 3D-ring cavities.} \textbf{a}: Polar plots of far-field intensity and polarization evolution for $q=2$-modes. \textbf{b}: Polarization states at intensity maxima of the upper and lower lobes for different inclination angles $\gamma=0^\circ,2^\circ,4^\circ,5^\circ.6^\circ,8^\circ,10^\circ,12^\circ,15^\circ$. \textbf{c}: Examples for electric field distributions inside ring resonators with inclination angles $\gamma = 0^\circ, 5^\circ, 10^\circ, 15^\circ$; the left panels are for $\gamma=15^\circ$. }
	\label{fig:f3}
\end{figure}

We shall now investigate the role of inclined resonator walls in order to understand  3D systems like cone-like tube-cavities or realistic (imperfect) 3D microresonators. The inclination angle $\gamma$ of the resonator wall with respect to the $z$-axis is illustrated in the inset of Fig.~\ref{fig:f3} \textbf{a}. Choosing $\gamma>0^\circ$ breaks the symmetry with respect to the $z=0$-plane. As a consequence, we expect that the modes inside the resonator and the far-field lobes will display an asymmetry as well. Analyzing this behaviour is crucial for all applications where the far field is taken as the sensing signal. \\
In the following, we will study this key question focussing on $q=2$-modes.
The upper row in Fig.~\ref{fig:f3} \textbf{a} shows the polar far-field plots for three inclination angles $\gamma=5^\circ,10^\circ,15^\circ$. The lower row shows the corresponding Poincar\'{e}-spheres of polarization displaying all observable polarization states when scanning the far field through the color-encoded $\theta$-range from $\theta_{\text{min}}=-33^\circ$ to $\theta_{\text{max}}=+33^\circ$.\\
A small wall inclination of $\gamma=5^\circ$ causes a rather slight asymmetry in the maximum intensity of the lobes while the shape of the far-field lobes is maintained and polarization orientation is only slightly tilted (shifted along the equator of the Poincar\'{e}) sphere, cf. Fig.~\ref{fig:f2} \textbf{b}). 
However, for higher inclination angles the far-field intensity and the shape of the polarization evolution change strongly both in terms of visible asymmetries in the far-field lobes and evolution on the Poincar\'{e} sphere.\\
Fig.~\ref{fig:f3} \textbf{b} shows the Poincar\'{e}-sphere of the polarization states at the far-field intensity maxima for different inclination angles $\gamma$ between $0^\circ$ and $15^\circ$. The red (blue) dots correspond to the upper (lower) lobes. The points right above and below $\textbf{L}_\text{V}$ correspond to $\gamma=0^\circ$. Via points at $\gamma=2^\circ,4^\circ,5^\circ,6^\circ,8^\circ,10^\circ,12^\circ$ the end point at $\gamma=15^\circ$, indicated by the polarization ellipse insets, is reached. Interestingly, the upper and lower lobes experience a different polarization evolution. The polarization states of the upper lobe remain elliptically polarized but experience a strong orientation tilt. In contrast, the polarization states of the lower lobe evolve from slightly elliptical to left-hand circular polarization.\\ 
The tilt of the polarization orientation can be explained and deduced from the vector diffraction model introduced in Eq.~(\ref{eq_kirchhoff_diff}) where we treat the wall of the resonator as an aperture that diffracts the waves inside the resonator into the far field. See section \textbf{I.B.} of the SM for more details and derivation of the following formulas. The far-field components $E_\varphi$ and $E_\theta$ in the case of cone-like 3D ring cavities can be written as
\begin{align}
E_\varphi &={} \cos\gamma~E_{\varphi,\text{ring}} + i\sin\gamma~E_{\varphi,\text{prec}} \label{equ:Ephi_cone} \\[5pt] 
E_\theta &={}  \cos\gamma~E_{\theta,\text{ring}} +  i\sin\gamma~E_{\theta,\text{prec}} \:, \label{equ:Ethe_cone} 
\end{align}  
where $E_{\varphi,\text{ring}}$ and $E_{\theta,\text{ring}}$ represent the far-field components that arise from the diffraction of a 3D ring structure, cf. Eq.~(23) of the SM and Eqs.~(24)-(27) of the SM. The additional terms that include  $E_{\varphi,\text{prec}}$ and $E_{\theta,\text{prec}}$ exist only for $|\gamma|>0$ (conical cavities) and arise from the precession of the electric field along its trajectory around the cone axis, cf. Eq.~(23) of the SM. The precession terms are phase shifted by $\pi/2$ w.r.t.  the ring diffraction terms as indicated by the prefactor $i$. As a consequence, both components $E_\varphi$ and $E_\theta$ undergo a phase change which increases with increasing inclination angle $\gamma$. This is the very origin of the phase $\delta$ between the far-field components $E_\varphi$ and $E_\theta$ which in turn results in a change of the orientation angle $\Psi$, cf. Eq.~(\ref{equ:def_Psi}). This explains the general feature of the increasing tilt angle $\Psi$ with increasing inclination angle $\gamma$ as shown in Fig.~\ref{fig:f3} \textbf{b}. \\ 
The different evolution of the upper and lower lobes can be explained by an asymmetric distortion of the amplitudes. By inclining the resonator wall, the mirror symmetry at the $z=0$ - plane is broken. Thus, we expect that the electric field amplitudes $\tilde{A}_\parallel(u')$ and $\tilde{A}_\phi(u')$ inside the resonator wall undergo a distortion. The left panels of Fig.~\ref{fig:f3} \textbf{c} show an example of such a distorted electric field. The panels on the right-hand side of Fig.~\ref{fig:f3} \textbf{c} display a comparison of $E_\parallel$ and $E_\phi$ at for different inclination angles $\gamma$ where $E_\parallel$ and $E_\phi$ correspond to the field distribution along the $z'$-axis (at half wall thickness) and the longitudinal $E_\phi$-component as indicated in the panels on the left-hand side.\\
First of all, we notice that the entire distributions of $E_\parallel$ and $E_\phi$ shift more towards the broader end of the cone (into the negative $z'$-direction) with increasing $\gamma$. Using the graph of $E_\parallel$ at $\gamma=15^\circ$ as an example, we see (i) that $E_\parallel$ for $z'>0$ decays to zero inside the cavity and (ii) that the distance between the maximum and minimum is compared to $\gamma=0^\circ$. These facts indicate that the  confinement length scale along the height of the resonator wall is reduced. As a result, the longitudinal component $E_\phi$ increases because of $E_\phi\propto\partial E_\parallel/\partial z'$. The increase of $E_\phi$ at $\gamma=15^\circ$ is noticeable through the increase of the height of the central maximum as indicated by the black arrow. Thus, the overall spin-orbit interaction is enhanced and explains the generally increasing ellipticity $\chi$ with increasing $\gamma$.\\
Concerning the lobe asymmetry, we use again the graph of $E_\parallel$ at $\gamma=15^\circ$ as an example. We see that the maximum ($z'>0$) shifts by an larger distance than the minimum ($z'<0$) as indicated by the black arrows. As a result, the upper part of the distribution is stretched while the lower part is compressed (asymmetric distortion). The stretching extends the confinement length scale locally and therefore, the spin-orbit coupling is reduced within this length scale. On the other hand, the compression reduces the confinement length scale locally and therefore, the spin-orbit coupling is enhanced within this length scale. This allows us to qualitatively explain the behaviour of the far-field polarization states of the two different lobes shown in Fig.~\ref{fig:f3} \textbf{b}.  As a result, the lower far-field lobes (blue points) emerging from the lower part of the distorted field distribution show an increased ellipticity with respect to the upper far-field lobes (red points) that emerge from the upper part of the distorted field distribution. \\
The property of asymmetrical distortion of the amplitudes is linked to the far-field components $E_\varphi$ and $E_\theta$ via the diffraction integral. We exemplarily show this for $E_{\varphi,\text{ring}}$. According to the derivations in the SM (cf. Eq.~(24) in SM), $E_{\varphi,\text{ring}}$ is given by:
\begin{align}
E_{\varphi,\text{ring}} = \left(K_\text{x,ring}\cos\varphi + K_\text{y,ring}\sin\varphi \right)\sin\theta - K_\text{z,ring}\cos\theta \:,
\end{align}
where $K_\text{x,ring}$, $K_\text{y,ring}$ and $K_\text{z,ring}$ are the components of the ring diffraction integral. For example, $K_\text{x,ring}$ is given by:
\begin{align}
K_\text{x,ring} &={} h R  \int_{-1/2}^{1/2}  \cos\gamma~\tilde{A}_{\parallel}(u')F_\text{x}(u')\exp{\left(-i\cos\gamma\tilde{k}_2u'\right)}  \d u' \:,
\end{align} 
where $F_\text{x}(u')$ is a function resulting from the $\phi'$ integration, cf. Eq.~(39) of the SM. Note that the $u'$-integration can not be treated approximately as a mere Fourier-Transform of the amplitudes as in the case of the 3D-ring, cf. Eq.~(\ref{equ:AzFF}). \\
We point out that Eqs.~(\ref{equ:Ephi_cone}) and (\ref{equ:Ethe_cone}) provide a formula which takes the effects of diffraction and  precession of the electric field into account. These effects determine the orientation angle $\Psi$ and the ellipticity $\chi$, cf. Eqs.~(\ref{equ:def_Psi}) and (\ref{equ:def_chi}). Especially important is that $\Psi$, $\chi$ and the handedness $\sigma$ depend on the far-field observation direction which is described by the angles $\varphi$ and $\theta$. We would like to highlight that the quantities $\Psi$ and $\chi$ are observable in the far field, but the WGM inside the resonator wall has a different and complicated polarization state. The connection between the polarization state of the WGM and the observable far-field quantities $(\Psi,\chi)$ is provided by diffraction and precession. \\
In~\cite{spinorbit-cones_2016}, an inclination of the far-field polarization orientation and an increase of the ellipticity was experimentally observed for inhomogeneous anisotropic cone-like cavities and explained in terms of non-cyclic geometric phases. Here, we find a similar behavior in the generic case of homogeneous isotropic cavities and an explanation in terms of diffraction theory.\\
\section{Complex cone-like tubular cavities}
\begin{figure}
	\centering
	\includegraphics[width=8.5cm]{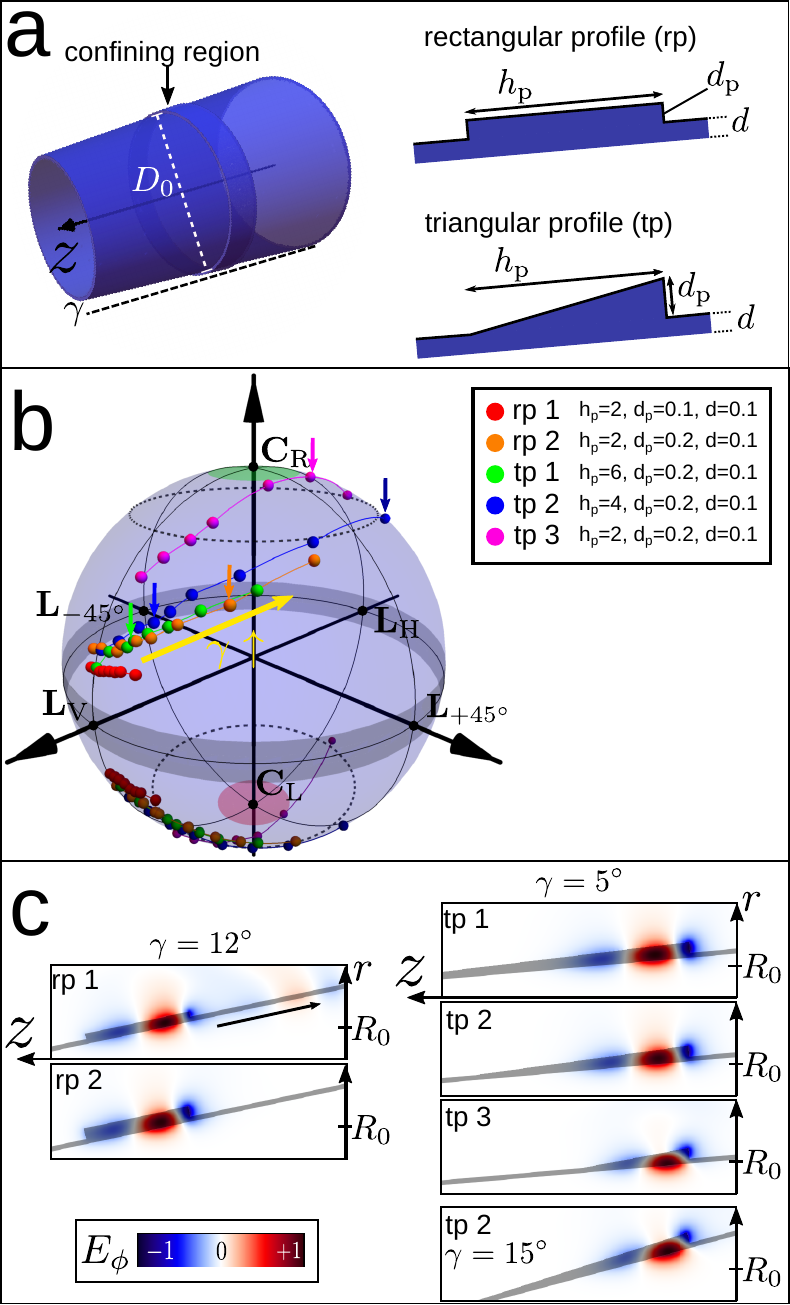}
	\caption{\textbf{Geometry and polarization states of cone-like tube-cavities.} \textbf{a}: Illustration of the geometry of the tube-cavity and different confinement profiles of the confining region. \textbf{b}: Polarization states at maximum far-field intensity of $q=2$-modes of different confinement profiles. The parameters displayed in the plot legend are given in $\mu m$. \textbf{c}: Comparison of field distributions $E_\phi$ at different profile configurations and inclination angles. See text for details. }
	\label{fig:f4}
\end{figure}

We have seen in the previous sections that the spin-orbit interaction of light depends sensitively on the confinement of light that is determined both by the resonator geometry and the specific resonance pattern. To further illustrate this intricate interplay,  we will now study complex, composite 3D cone-like tubular cavities where the confining region of the cavity is extended by an additional layer of cavity material, see Fig.~\ref{fig:f4} \textbf{a}. This resonator wall geometry is inspired by rolled-up cones where regions of different wall thicknesses  emerge, e.g. cf.~\cite{wgm-rings_2007,Belanos_2012,spinorbit-cones_2016} and by cavities where an extra layer of material results from an etching process, cf.~\cite{wgm-bottles_2012}. \\
\noindent The confining region is realized by two different axial profiles, a rectangular profile (rp) and triangular profile (tp), as shown on the right-hand side of Fig.~\ref{fig:f4} \textbf{a}.  We consider cone-like cavities with a mean diameter $D_0=6\mu m$ and total height $h_\text{T}=20\mu m$. We will investigate the far-field polarization states and how they depend on (i) the confining profile and (ii) the inclination angle $\gamma$ of the tube wall. The results are shown in Fig.~\ref{fig:f4} \textbf{b}. The far-field polarization states of maximum intensity of $q=2$-modes confined in 5 different axial profiles are shown on the Poincar\'{e} sphere with its upper (lower) hemisphere corresponding to the upper (lower) lobes. The yellow arrow indicates the generic polarization evolution for increasing inclination angle $\gamma$ (again, with the data points closest to the longitude of $\textbf{L}_\text{V}$ corresponding to $\gamma=0^\circ$). We observe the general feature that with increasing $\gamma$ the polarization states evolve into the direction of $\textbf{L}_{+45^\circ}$ while getting closer to the poles of the sphere. In other words, orientation tilt $\Psi$ and the ellipticity $\chi$ increase with increasing wall inclination. This behaviour is very similar to the one observed before, cf. Fig.~\ref{fig:f3} \textbf{b}.  
The evolution of the polarization state depends on the 
confining profile that in turn controls the mode's axial field distribution, 
and hence, following Eq.~(\ref{equ:Ez_beam}), the spin-orbit interaction. In case of the rectangular profile, the far-field polarization states depend strongly on the thickness of the profile $d_\text{p}$ as represented by the red (rp 1) and orange (rp 2) data points on the Poincar\'{e} sphere in Fig.~\ref{fig:f4} \textbf{b}. The inclination angles of the rectangular profile 1 range from $\gamma_\text{min}=0^\circ$ up to $\gamma_\text{max}=8^\circ$ in steps of one degree. For higher angles than $\gamma_\text{max}$ the modes become unstable and propagate downwards (towards the broader end of) the cone because the confining profile is too thin to stably host a WGM resonance. This finding is illustrated on the left-hand side of Fig.~\ref{fig:f4} \textbf{c}. The upper (lower) image shows a section around the confining region rp 1 (rp 2) for  $\gamma=12^\circ$, overlaid with the electric field component $E_\phi$ obtained from FDTD simulations. 
We clearly see that a wave packet is leaving the profile rp 1 and propagating towards the broader part of the cone (black arrow). This indicates an unstable mode and no stationary far-field polarization states can be found. Contrary to this, the thicker profile rp 2 ensures the existence of a stable WGM-type resonance, see the lower left panel in Fig.~\ref{fig:f4}c that corresponds to the state marked by an orange arrow on the Poincar\'{e} sphere in Fig.~\ref{fig:f4} \textbf{b}.\\
The upper three images on the right-hand side of Fig.~\ref{fig:f4} \textbf{c} compare the electric field distribution $E_\phi$ at different configurations of the triangular profile tp 1, tp 2 and tp 3 but all at constant wall inclination angle $\gamma=5^\circ$. The corresponding far-field polarization states are marked by a green (tp 1), blue (tp 2)  and magenta (tp 3) arrow on the 
Poincar\'{e} sphere of Fig.~\ref{fig:f4} \textbf{b}. We see that the difference between the field distribution in tp 1 and tp 2 is small. As a consequence the corresponding far-field polarization states are located in close proximity on the Poincar\'{e} sphere, cf. Fig.~\ref{fig:f4} \textbf{b}. The highlighted polarization state corresponding to tp 2 shows a slightly higher tilt and ellipticity because the profile tp 2 provides stronger confinement, and hence stronger spin-orbit interaction due to the reduced profile height $h_\text{p}$.\\
On the other hand, the difference between the field distributions in tp 2 and tp 3 is more pronounced due to the further reduced profile height of tp 3. The $E_\phi$ distribution is now visibly characterized by a strong mode distortion 
that causes a higher spin-orbit interaction.  As a result, the corresponding far-field polarization state of tp 3 (highlighted by the magenta arrow on the Poincar\'{e} sphere in Fig.~\ref{fig:f4} \textbf{b}) is almost circularly polarized.\\
For comparison, the lowest panel on the right-hand side of Fig.~\ref{fig:f4} \textbf{c} shows the $E_\phi$ distribution of tp 2 at the higher inclination angle of $\gamma=15^\circ$. Interestingly, this $E_\phi$ distribution is very similar to that of tp 3 because it displays a similar distortion, confer the red areas of the $E_\phi$ distributions. We conclude that the high inclination angle of $\gamma=15^\circ$ of tp 2 results in a comparable confinement and distortion of the mode as caused by the profile tp 3 at lower angle. As a result the corresponding far-field polarization states display similar features indicated by the dark blue and magenta arrow on the Poincar\'{e} sphere.\\
The overall increasing tilt of the polarization orientation and ellipticity caused by the triangular profiles result from a enhanced precession of the electric field and a stronger spin-orbit coupling, respectively. The stronger spin-orbit coupling can be explained by an increased axial confinement. The enhanced precession of the electric field can be explained by an increased effective inclination angle $\gamma_\text{eff}$. In addition to the inclined cone wall described by $\gamma$, the triangular profile provides a further local inclination given by $\tan\gamma_\text{p}=d_\text{p}/h_\text{p}$. The interplay of both inclinations leads to $\gamma_\text{eff}>\gamma$ and thus, 
according to the introduced diffraction model, an increased inclination angle leads to an increased tilt of the polarization orientation.\\
\section{Summary}
\noindent We have performed FDTD simulations in order to investigate the spin-orbit coupling of light in three-dimensional cylindrical and tube-like whispering gallery mode resonators. \\
We have shown that the spin-orbit interaction in cylindrical ring cavities results in elliptical far-field polarization and spatial separation of left and right elliptically polarized light in the far field but without tilting of the orientation angle. The ellipticity and spatial polarization separation of the far field is more pronounced in axially higher excited whispering gallery modes due to increased spin-orbit coupling.\\
Furthermore, we have shown that the inclination of the resonator wall realized by cylindrical ring-like cone-cavities enhances the ellipticity of the far field further and induces a tilt of the far-field polarization orientation even for homogeneous and isotropic material systems. The enhancement of the far-field ellipticity arises from the asymmetric distortion of the electric field distribution at the resonator wall. This asymmetric distortion causes different elliptical polarization states of upper and lower far-field lobes. The tilt of the polarization orientation arises from the precession of the electric field vector along its trajectory around the cone axis. The connection between the local polarization state of the whispering gallery mode inside the resonator and the observable far-field quantities $(\Psi,\chi)$ is provided by the introduced diffraction model. We emphasize that the polarization orientation $\Psi$, the ellipticity $\chi$ and the handedness $\sigma$ depend on the direction of observation described by the far-field angles $(\varphi,\theta)$.\\
In the end, we have investigated complex cone-like tubular cavities with different confining profiles. Similar to the case of ring-like cone-cavities, the tilt of the polarization orientation $\Psi$ and the ellipticity $\chi$  increase with increasing wall inclination $\gamma$. Furthermore, we have discovered that a triangular profile enhances the tilt of the orientation and the ellipticity because of an increased effective inclination $\gamma_\text{eff}$ and a stronger spin-orbit coupling, respectively.\\
Our results demonstrate the importance and variety of spin-orbit coupling of light in three-dimensional whispering gallery mode resonators as a fundamental effect and may be important for optical information technology or polarization dependent sensing applications.\\

\noindent\textbf{FDTD-Method}\\
All data presented in this work was obtained from FDTD-calculation based on the open source software package MEEP~\cite{MEEP}.\\
We give a short description of the procedure: In the first step, we have computed the resonance frequencies of TE-like modes for the following structures: 3D-ring cavities (cf. section II) , cone-like ring cavities (cf. section III) and cone-like tubular cavities (cf. section IV).\\
In the second step, we have computed the electric field distribution of the resonances and let the fields evolve for one period in order to make 20 snapshots of the electric field in the region around the cavity (near field).\\ 
In addition to this, we computed 20 snapshots of the electric field on a circular arc with radius of $r_\text{FF}=50\mu\text{m}$ (approximately satisfying the far-field condition $r_\text{FF}\gg R^2/\lambda_{q=1}$$\approx 6\mu\text{m}$ ) within the range of $\theta=\pm45^\circ$ in steps of $\Delta\theta=1^\circ$. These far-field snapshots were used to calculate $|E_\text{FF}|^2$ (as a measure of the far-field intensity) and the orientation angle $\Psi$ and the ellipticity angle $\chi$ of the polarization ellipse. The handedness $\sigma$ was determined by the rotation sense of the electric field vector describing the polarization ellipse. Please note that due to the cylindrical symmetry of the considered structures we have used cylindrical coordinates in MEEP. As a result, the computed fields have the form $E=f(r,z)\exp{(i m \phi)}$ and thus they represent azimuthally propagating fields.\\    

\begin{acknowledgments}
	This work was partly supported by Emmy-Noether programme of the German Research Foundation (DFG).
\end{acknowledgments}

\bibliography{references.bib}

\begin{thebibliography}{46}%
\makeatletter
\providecommand \@ifxundefined [1]{%
 \@ifx{#1\undefined}
}%
\providecommand \@ifnum [1]{%
 \ifnum #1\expandafter \@firstoftwo
 \else \expandafter \@secondoftwo
 \fi
}%
\providecommand \@ifx [1]{%
 \ifx #1\expandafter \@firstoftwo
 \else \expandafter \@secondoftwo
 \fi
}%
\providecommand \natexlab [1]{#1}%
\providecommand \enquote  [1]{``#1''}%
\providecommand \bibnamefont  [1]{#1}%
\providecommand \bibfnamefont [1]{#1}%
\providecommand \citenamefont [1]{#1}%
\providecommand \href@noop [0]{\@secondoftwo}%
\providecommand \href [0]{\begingroup \@sanitize@url \@href}%
\providecommand \@href[1]{\@@startlink{#1}\@@href}%
\providecommand \@@href[1]{\endgroup#1\@@endlink}%
\providecommand \@sanitize@url [0]{\catcode `\\12\catcode `\$12\catcode
  `\&12\catcode `\#12\catcode `\^12\catcode `\_12\catcode `\%12\relax}%
\providecommand \@@startlink[1]{}%
\providecommand \@@endlink[0]{}%
\providecommand \url  [0]{\begingroup\@sanitize@url \@url }%
\providecommand \@url [1]{\endgroup\@href {#1}{\urlprefix }}%
\providecommand \urlprefix  [0]{URL }%
\providecommand \Eprint [0]{\href }%
\providecommand \doibase [0]{http://dx.doi.org/}%
\providecommand \selectlanguage [0]{\@gobble}%
\providecommand \bibinfo  [0]{\@secondoftwo}%
\providecommand \bibfield  [0]{\@secondoftwo}%
\providecommand \translation [1]{[#1]}%
\providecommand \BibitemOpen [0]{}%
\providecommand \bibitemStop [0]{}%
\providecommand \bibitemNoStop [0]{.\EOS\space}%
\providecommand \EOS [0]{\spacefactor3000\relax}%
\providecommand \BibitemShut  [1]{\csname bibitem#1\endcsname}%
\let\auto@bib@innerbib\@empty
\bibitem [{\citenamefont {Yamamoto}\ and\ \citenamefont
  {Slusher}(1993)}]{Chang1996}%
  \BibitemOpen
  \bibfield  {author} {\bibinfo {author} {\bibfnamefont {Y.}~\bibnamefont
  {Yamamoto}}\ and\ \bibinfo {author} {\bibfnamefont {R.~E.}\ \bibnamefont
  {Slusher}},\ }\href {\doibase 10.1063/1.881356} {\emph {\bibinfo {title}
  {Physics Today}}},\ edited by\ \bibinfo {editor} {\bibfnamefont {A.~J.~C.}\
  \bibnamefont {{R. K. Chang}}},\ Vol.~\bibinfo {volume} {46}\ (\bibinfo
  {publisher} {World Scientific, Singapore},\ \bibinfo {year} {1993})\ pp.\
  \bibinfo {pages} {66--73}\BibitemShut {NoStop}%
\bibitem [{\citenamefont {Vahala}(2003)}]{Vahala2003}%
  \BibitemOpen
  \bibfield  {author} {\bibinfo {author} {\bibfnamefont {K.~J.}\ \bibnamefont
  {Vahala}},\ }\href {http://dx.doi.org/10.1038/nature01939} {\bibfield
  {journal} {\bibinfo  {journal} {Nature}\ }\textbf {\bibinfo {volume} {424}},\
  \bibinfo {pages} {839} (\bibinfo {year} {2003})}\BibitemShut {NoStop}%
\bibitem [{\citenamefont {Collot}\ \emph {et~al.}(1993)\citenamefont {Collot},
  \citenamefont {Lef{\`{e}}vre-Seguin}, \citenamefont {Brune}, \citenamefont
  {Raimond},\ and\ \citenamefont {Haroche}}]{Collot1993}%
  \BibitemOpen
  \bibfield  {author} {\bibinfo {author} {\bibfnamefont {L.}~\bibnamefont
  {Collot}}, \bibinfo {author} {\bibfnamefont {V.}~\bibnamefont
  {Lef{\`{e}}vre-Seguin}}, \bibinfo {author} {\bibfnamefont {M.}~\bibnamefont
  {Brune}}, \bibinfo {author} {\bibfnamefont {J.~M.}\ \bibnamefont {Raimond}},
  \ and\ \bibinfo {author} {\bibfnamefont {S.}~\bibnamefont {Haroche}},\ }\href
  {http://stacks.iop.org/0295-5075/23/i=5/a=005} {\bibfield  {journal}
  {\bibinfo  {journal} {EPL}\ }\textbf {\bibinfo {volume} {23}},\ \bibinfo
  {pages} {327} (\bibinfo {year} {1993})}\BibitemShut {NoStop}%
\bibitem [{\citenamefont {Gorodetsky}\ \emph {et~al.}(2000)\citenamefont
  {Gorodetsky}, \citenamefont {Pryamikov},\ and\ \citenamefont
  {Ilchenko}}]{Gorodetsky2000}%
  \BibitemOpen
  \bibfield  {author} {\bibinfo {author} {\bibfnamefont {M.~L.}\ \bibnamefont
  {Gorodetsky}}, \bibinfo {author} {\bibfnamefont {A.~D.}\ \bibnamefont
  {Pryamikov}}, \ and\ \bibinfo {author} {\bibfnamefont {V.~S.}\ \bibnamefont
  {Ilchenko}},\ }\href {\doibase 10.1364/JOSAB.17.001051} {\bibfield  {journal}
  {\bibinfo  {journal} {J. Opt. Soc. Am. B}\ }\textbf {\bibinfo {volume}
  {17}},\ \bibinfo {pages} {1051} (\bibinfo {year} {2000})}\BibitemShut
  {NoStop}%
\bibitem [{\citenamefont {Vollmer}\ \emph {et~al.}(2012)\citenamefont
  {Vollmer}, \citenamefont {Yang},\ and\ \citenamefont
  {Fainman}}]{VollmerYang2012}%
  \BibitemOpen
  \bibfield  {author} {\bibinfo {author} {\bibfnamefont {F.}~\bibnamefont
  {Vollmer}}, \bibinfo {author} {\bibfnamefont {L.}~\bibnamefont {Yang}}, \
  and\ \bibinfo {author} {\bibfnamefont {S.}~\bibnamefont {Fainman}},\ }\href
  {\doibase 10.1515/nanoph-2012-0021} {\bibfield  {journal} {\bibinfo
  {journal} {Nanophotonics}\ }\textbf {\bibinfo {volume} {1}},\ \bibinfo
  {pages} {267} (\bibinfo {year} {2012})}\BibitemShut {NoStop}%
\bibitem [{\citenamefont {Ilchenko}\ \emph {et~al.}(2001)\citenamefont
  {Ilchenko}, \citenamefont {Gorodetsky}, \citenamefont {Yao},\ and\
  \citenamefont {Maleki}}]{Ilchenko2001}%
  \BibitemOpen
  \bibfield  {author} {\bibinfo {author} {\bibfnamefont {V.~S.}\ \bibnamefont
  {Ilchenko}}, \bibinfo {author} {\bibfnamefont {M.~L.}\ \bibnamefont
  {Gorodetsky}}, \bibinfo {author} {\bibfnamefont {X.~S.}\ \bibnamefont {Yao}},
  \ and\ \bibinfo {author} {\bibfnamefont {L.}~\bibnamefont {Maleki}},\ }\href
  {\doibase 10.1364/OL.26.000256} {\bibfield  {journal} {\bibinfo  {journal}
  {Opt. Lett.}\ }\textbf {\bibinfo {volume} {26}},\ \bibinfo {pages} {256}
  (\bibinfo {year} {2001})}\BibitemShut {NoStop}%
\bibitem [{\citenamefont {Armani}\ \emph {et~al.}(2003)\citenamefont {Armani},
  \citenamefont {Kippenberg}, \citenamefont {Spillane},\ and\ \citenamefont
  {Vahala}}]{Armani2003}%
  \BibitemOpen
  \bibfield  {author} {\bibinfo {author} {\bibfnamefont {D.~K.}\ \bibnamefont
  {Armani}}, \bibinfo {author} {\bibfnamefont {T.~J.}\ \bibnamefont
  {Kippenberg}}, \bibinfo {author} {\bibfnamefont {S.~M.}\ \bibnamefont
  {Spillane}}, \ and\ \bibinfo {author} {\bibfnamefont {K.~J.}\ \bibnamefont
  {Vahala}},\ }\href {http://dx.doi.org/10.1038/nature01371} {\bibfield
  {journal} {\bibinfo  {journal} {Nature}\ }\textbf {\bibinfo {volume} {421}},\
  \bibinfo {pages} {925} (\bibinfo {year} {2003})}\BibitemShut {NoStop}%
\bibitem [{\citenamefont {McCall}\ \emph {et~al.}(1992)\citenamefont {McCall},
  \citenamefont {Levi}, \citenamefont {Slusher}, \citenamefont {Pearton},\ and\
  \citenamefont {Logan}}]{McCall1992}%
  \BibitemOpen
  \bibfield  {author} {\bibinfo {author} {\bibfnamefont {S.~L.}\ \bibnamefont
  {McCall}}, \bibinfo {author} {\bibfnamefont {A.~F.~J.}\ \bibnamefont {Levi}},
  \bibinfo {author} {\bibfnamefont {R.~E.}\ \bibnamefont {Slusher}}, \bibinfo
  {author} {\bibfnamefont {S.~J.}\ \bibnamefont {Pearton}}, \ and\ \bibinfo
  {author} {\bibfnamefont {R.~A.}\ \bibnamefont {Logan}},\ }\href {\doibase
  http://dx.doi.org/10.1063/1.106688} {\bibfield  {journal} {\bibinfo
  {journal} {Appl. Phys. Lett.}\ }\textbf {\bibinfo {volume} {60}},\ \bibinfo
  {pages} {289} (\bibinfo {year} {1992})}\BibitemShut {NoStop}%
\bibitem [{\citenamefont {Michael}\ \emph {et~al.}(2007)\citenamefont
  {Michael}, \citenamefont {Srinivasan}, \citenamefont {Johnson}, \citenamefont
  {Painter}, \citenamefont {Lee}, \citenamefont {Hennessy}, \citenamefont
  {Kim},\ and\ \citenamefont {Hu}}]{Michael2007}%
  \BibitemOpen
  \bibfield  {author} {\bibinfo {author} {\bibfnamefont {C.~P.}\ \bibnamefont
  {Michael}}, \bibinfo {author} {\bibfnamefont {K.}~\bibnamefont {Srinivasan}},
  \bibinfo {author} {\bibfnamefont {T.~J.}\ \bibnamefont {Johnson}}, \bibinfo
  {author} {\bibfnamefont {O.}~\bibnamefont {Painter}}, \bibinfo {author}
  {\bibfnamefont {K.~H.}\ \bibnamefont {Lee}}, \bibinfo {author} {\bibfnamefont
  {K.}~\bibnamefont {Hennessy}}, \bibinfo {author} {\bibfnamefont
  {H.}~\bibnamefont {Kim}}, \ and\ \bibinfo {author} {\bibfnamefont
  {E.}~\bibnamefont {Hu}},\ }\href {\doibase
  http://dx.doi.org/10.1063/1.2435608} {\bibfield  {journal} {\bibinfo
  {journal} {Appl. Phys. Lett.}\ }\textbf {\bibinfo {volume} {90}},\  (\bibinfo
  {year} {2007})}\BibitemShut {NoStop}%
\bibitem [{\citenamefont {Sumetsky}(2004)}]{wgm_bottles2004}%
  \BibitemOpen
  \bibfield  {author} {\bibinfo {author} {\bibfnamefont {M.}~\bibnamefont
  {Sumetsky}},\ }\href {\doibase 10.1364/ol.29.000008} {\bibfield  {journal}
  {\bibinfo  {journal} {Optics Letters}\ }\textbf {\bibinfo {volume} {29}},\
  \bibinfo {pages} {8} (\bibinfo {year} {2004})}\BibitemShut {NoStop}%
\bibitem [{\citenamefont {Strelow}\ \emph {et~al.}(2007)\citenamefont
  {Strelow}, \citenamefont {Schultz}, \citenamefont {Rehberg}, \citenamefont
  {Welsch}, \citenamefont {Heyn}, \citenamefont {Heitmann},\ and\ \citenamefont
  {Kipp}}]{wgm-rings_2007}%
  \BibitemOpen
  \bibfield  {author} {\bibinfo {author} {\bibfnamefont {C.}~\bibnamefont
  {Strelow}}, \bibinfo {author} {\bibfnamefont {C.~M.}\ \bibnamefont
  {Schultz}}, \bibinfo {author} {\bibfnamefont {H.}~\bibnamefont {Rehberg}},
  \bibinfo {author} {\bibfnamefont {H.}~\bibnamefont {Welsch}}, \bibinfo
  {author} {\bibfnamefont {C.}~\bibnamefont {Heyn}}, \bibinfo {author}
  {\bibfnamefont {D.}~\bibnamefont {Heitmann}}, \ and\ \bibinfo {author}
  {\bibfnamefont {T.}~\bibnamefont {Kipp}},\ }\href {\doibase
  10.1103/PhysRevB.76.045303} {\bibfield  {journal} {\bibinfo  {journal}
  {Physical Review B - Condensed Matter and Materials Physics}\ }\textbf
  {\bibinfo {volume} {76}},\ \bibinfo {pages} {1} (\bibinfo {year}
  {2007})}\BibitemShut {NoStop}%
\bibitem [{\citenamefont {Strelow}\ \emph {et~al.}(2008)\citenamefont
  {Strelow}, \citenamefont {Rehberg}, \citenamefont {Schultz}, \citenamefont
  {Welsch}, \citenamefont {Heyn}, \citenamefont {Heitmann},\ and\ \citenamefont
  {Kipp}}]{wgm-bottles_2008}%
  \BibitemOpen
  \bibfield  {author} {\bibinfo {author} {\bibfnamefont {C.}~\bibnamefont
  {Strelow}}, \bibinfo {author} {\bibfnamefont {H.}~\bibnamefont {Rehberg}},
  \bibinfo {author} {\bibfnamefont {C.~M.}\ \bibnamefont {Schultz}}, \bibinfo
  {author} {\bibfnamefont {H.}~\bibnamefont {Welsch}}, \bibinfo {author}
  {\bibfnamefont {C.}~\bibnamefont {Heyn}}, \bibinfo {author} {\bibfnamefont
  {D.}~\bibnamefont {Heitmann}}, \ and\ \bibinfo {author} {\bibfnamefont
  {T.}~\bibnamefont {Kipp}},\ }\href {\doibase 10.1103/PhysRevLett.101.127403}
  {\bibfield  {journal} {\bibinfo  {journal} {Physical Review Letters}\
  }\textbf {\bibinfo {volume} {101}},\ \bibinfo {pages} {1} (\bibinfo {year}
  {2008})}\BibitemShut {NoStop}%
\bibitem [{\citenamefont {Li}\ \emph {et~al.}(2009)\citenamefont {Li},
  \citenamefont {Vicknesh},\ and\ \citenamefont {Mi}}]{wgm-bottles_2009}%
  \BibitemOpen
  \bibfield  {author} {\bibinfo {author} {\bibfnamefont {F.}~\bibnamefont
  {Li}}, \bibinfo {author} {\bibfnamefont {S.}~\bibnamefont {Vicknesh}}, \ and\
  \bibinfo {author} {\bibfnamefont {Z.}~\bibnamefont {Mi}},\ }\href {\doibase
  10.1049/el.2009.1076} {\bibfield  {journal} {\bibinfo  {journal} {Electronics
  Letters}\ }\textbf {\bibinfo {volume} {45}},\ \bibinfo {pages} {645}
  (\bibinfo {year} {2009})}\BibitemShut {NoStop}%
\bibitem [{\citenamefont {Strelow}\ \emph {et~al.}(2012)\citenamefont
  {Strelow}, \citenamefont {Schultz}, \citenamefont {Rehberg}, \citenamefont
  {Sauer}, \citenamefont {Welsch}, \citenamefont {Stemmann}, \citenamefont
  {Heyn}, \citenamefont {Heitmann},\ and\ \citenamefont
  {Kipp}}]{wgm-bottles_2012}%
  \BibitemOpen
  \bibfield  {author} {\bibinfo {author} {\bibfnamefont {C.}~\bibnamefont
  {Strelow}}, \bibinfo {author} {\bibfnamefont {C.~M.}\ \bibnamefont
  {Schultz}}, \bibinfo {author} {\bibfnamefont {H.}~\bibnamefont {Rehberg}},
  \bibinfo {author} {\bibfnamefont {M.}~\bibnamefont {Sauer}}, \bibinfo
  {author} {\bibfnamefont {H.}~\bibnamefont {Welsch}}, \bibinfo {author}
  {\bibfnamefont {A.}~\bibnamefont {Stemmann}}, \bibinfo {author}
  {\bibfnamefont {C.}~\bibnamefont {Heyn}}, \bibinfo {author} {\bibfnamefont
  {D.}~\bibnamefont {Heitmann}}, \ and\ \bibinfo {author} {\bibfnamefont
  {T.}~\bibnamefont {Kipp}},\ }\href {\doibase 10.1103/PhysRevB.85.155329}
  {\bibfield  {journal} {\bibinfo  {journal} {Physical Review B - Condensed
  Matter and Materials Physics}\ }\textbf {\bibinfo {volume} {85}},\ \bibinfo
  {pages} {1} (\bibinfo {year} {2012})}\BibitemShut {NoStop}%
\bibitem [{\citenamefont {Ma}\ \emph {et~al.}(2016)\citenamefont {Ma},
  \citenamefont {Li}, \citenamefont {Fomin}, \citenamefont {Hentschel},
  \citenamefont {G{\"{o}}tte}, \citenamefont {Yin}, \citenamefont {Jorgensen},\
  and\ \citenamefont {Schmidt}}]{spinorbit-cones_2016}%
  \BibitemOpen
  \bibfield  {author} {\bibinfo {author} {\bibfnamefont {L.~B.}\ \bibnamefont
  {Ma}}, \bibinfo {author} {\bibfnamefont {S.~L.}\ \bibnamefont {Li}}, \bibinfo
  {author} {\bibfnamefont {V.~M.}\ \bibnamefont {Fomin}}, \bibinfo {author}
  {\bibfnamefont {M.}~\bibnamefont {Hentschel}}, \bibinfo {author}
  {\bibfnamefont {J.~B.}\ \bibnamefont {G{\"{o}}tte}}, \bibinfo {author}
  {\bibfnamefont {Y.}~\bibnamefont {Yin}}, \bibinfo {author} {\bibfnamefont
  {M.~R.}\ \bibnamefont {Jorgensen}}, \ and\ \bibinfo {author} {\bibfnamefont
  {O.~G.}\ \bibnamefont {Schmidt}},\ }\href {\doibase 10.1038/ncomms10983}
  {\bibfield  {journal} {\bibinfo  {journal} {Nature Communications}\ }\textbf
  {\bibinfo {volume} {7}},\ \bibinfo {pages} {4} (\bibinfo {year}
  {2016})}\BibitemShut {NoStop}%
\bibitem [{\citenamefont {Huang}\ and\ \citenamefont
  {Mei}(2017)}]{wgm-tubes_2017}%
  \BibitemOpen
  \bibfield  {author} {\bibinfo {author} {\bibfnamefont {G.}~\bibnamefont
  {Huang}}\ and\ \bibinfo {author} {\bibfnamefont {Y.}~\bibnamefont {Mei}},\
  }\href {\doibase 10.1039/c7tc00283a} {\bibfield  {journal} {\bibinfo
  {journal} {Journal of Materials Chemistry C}\ }\textbf {\bibinfo {volume}
  {5}},\ \bibinfo {pages} {2758} (\bibinfo {year} {2017})}\BibitemShut
  {NoStop}%
\bibitem [{\citenamefont {Wang}\ \emph {et~al.}(2019)\citenamefont {Wang},
  \citenamefont {Yin}, \citenamefont {Yang}, \citenamefont {Hao}, \citenamefont
  {Tang}, \citenamefont {Wang}, \citenamefont {Saggau}, \citenamefont
  {Karnaushenko}, \citenamefont {Yan}, \citenamefont {Huang}, \citenamefont
  {Ma},\ and\ \citenamefont {Schmidt}}]{wgm-tubes_2019}%
  \BibitemOpen
  \bibfield  {author} {\bibinfo {author} {\bibfnamefont {J.}~\bibnamefont
  {Wang}}, \bibinfo {author} {\bibfnamefont {Y.}~\bibnamefont {Yin}}, \bibinfo
  {author} {\bibfnamefont {Y.~D.}\ \bibnamefont {Yang}}, \bibinfo {author}
  {\bibfnamefont {Q.}~\bibnamefont {Hao}}, \bibinfo {author} {\bibfnamefont
  {M.}~\bibnamefont {Tang}}, \bibinfo {author} {\bibfnamefont {X.}~\bibnamefont
  {Wang}}, \bibinfo {author} {\bibfnamefont {C.~N.}\ \bibnamefont {Saggau}},
  \bibinfo {author} {\bibfnamefont {D.}~\bibnamefont {Karnaushenko}}, \bibinfo
  {author} {\bibfnamefont {X.}~\bibnamefont {Yan}}, \bibinfo {author}
  {\bibfnamefont {Y.~Z.}\ \bibnamefont {Huang}}, \bibinfo {author}
  {\bibfnamefont {L.}~\bibnamefont {Ma}}, \ and\ \bibinfo {author}
  {\bibfnamefont {O.~G.}\ \bibnamefont {Schmidt}},\ }\href {\doibase
  10.1021/acsphotonics.9b00992} {\bibfield  {journal} {\bibinfo  {journal} {ACS
  Photonics}\ }\textbf {\bibinfo {volume} {6}},\ \bibinfo {pages} {2537}
  (\bibinfo {year} {2019})}\BibitemShut {NoStop}%
\bibitem [{\citenamefont {Lacey}\ and\ \citenamefont
  {Wang}(2001)}]{Lacey_2001}%
  \BibitemOpen
  \bibfield  {author} {\bibinfo {author} {\bibfnamefont {S.}~\bibnamefont
  {Lacey}}\ and\ \bibinfo {author} {\bibfnamefont {H.}~\bibnamefont {Wang}},\
  }\href {\doibase 10.1364/OL.26.001943} {\bibfield  {journal} {\bibinfo
  {journal} {Optics Letters}\ }\textbf {\bibinfo {volume} {26}},\ \bibinfo
  {pages} {1943} (\bibinfo {year} {2001})}\BibitemShut {NoStop}%
\bibitem [{\citenamefont {Schwefel}\ \emph {et~al.}(2005)\citenamefont
  {Schwefel}, \citenamefont {Stone},\ and\ \citenamefont
  {Tureci}}]{Schwefel_2005}%
  \BibitemOpen
  \bibfield  {author} {\bibinfo {author} {\bibfnamefont {H.~G.~L.}\
  \bibnamefont {Schwefel}}, \bibinfo {author} {\bibfnamefont {A.~D.}\
  \bibnamefont {Stone}}, \ and\ \bibinfo {author} {\bibfnamefont {H.~E.}\
  \bibnamefont {Tureci}},\ }\href {\doibase 10.1364/JOSAB.22.002295} {\bibfield
   {journal} {\bibinfo  {journal} {Journal of the Optical Society of America
  B}\ }\textbf {\bibinfo {volume} {22}},\ \bibinfo {pages} {2295} (\bibinfo
  {year} {2005})}\BibitemShut {NoStop}%
\bibitem [{\citenamefont {Teraoka}\ and\ \citenamefont
  {Arnold}(2006)}]{Teraoka_2006}%
  \BibitemOpen
  \bibfield  {author} {\bibinfo {author} {\bibfnamefont {I.}~\bibnamefont
  {Teraoka}}\ and\ \bibinfo {author} {\bibfnamefont {S.}~\bibnamefont
  {Arnold}},\ }\href {\doibase 10.1364/JOSAB.23.001381} {\bibfield  {journal}
  {\bibinfo  {journal} {Journal of the Optical Society of America B}\ }\textbf
  {\bibinfo {volume} {23}},\ \bibinfo {pages} {1381} (\bibinfo {year}
  {2006})}\BibitemShut {NoStop}%
\bibitem [{\citenamefont {Kreismann}\ \emph {et~al.}(2017)\citenamefont
  {Kreismann}, \citenamefont {Sinzinger},\ and\ \citenamefont
  {Hentschel}}]{Kreismann2017}%
  \BibitemOpen
  \bibfield  {author} {\bibinfo {author} {\bibfnamefont {J.}~\bibnamefont
  {Kreismann}}, \bibinfo {author} {\bibfnamefont {S.}~\bibnamefont
  {Sinzinger}}, \ and\ \bibinfo {author} {\bibfnamefont {M.}~\bibnamefont
  {Hentschel}},\ }\href {\doibase 10.1103/PhysRevA.95.011801} {\bibfield
  {journal} {\bibinfo  {journal} {Physical Review A}\ }\textbf {\bibinfo
  {volume} {95}},\ \bibinfo {pages} {1} (\bibinfo {year} {2017})}\BibitemShut
  {NoStop}%
\bibitem [{\citenamefont {Kreismann}\ and\ \citenamefont
  {Hentschel}(2018)}]{Kreismann2018}%
  \BibitemOpen
  \bibfield  {author} {\bibinfo {author} {\bibfnamefont {J.}~\bibnamefont
  {Kreismann}}\ and\ \bibinfo {author} {\bibfnamefont {M.}~\bibnamefont
  {Hentschel}},\ }\href {\doibase 10.1209/0295-5075/121/24001} {\bibfield
  {journal} {\bibinfo  {journal} {EPL (Europhysics Letters)}\ }\textbf
  {\bibinfo {volume} {121}},\ \bibinfo {pages} {24001} (\bibinfo {year}
  {2018})}\BibitemShut {NoStop}%
\bibitem [{\citenamefont {Gladyshev}\ \emph {et~al.}(2018)\citenamefont
  {Gladyshev}, \citenamefont {Bogdanov}, \citenamefont {Kapitanova},
  \citenamefont {Rybin}, \citenamefont {Koshelev}, \citenamefont {Sadrieva},
  \citenamefont {Samusev}, \citenamefont {Kivshar},\ and\ \citenamefont
  {Limonov}}]{Gladyshev_2018}%
  \BibitemOpen
  \bibfield  {author} {\bibinfo {author} {\bibfnamefont {S.~A.}\ \bibnamefont
  {Gladyshev}}, \bibinfo {author} {\bibfnamefont {A.~A.}\ \bibnamefont
  {Bogdanov}}, \bibinfo {author} {\bibfnamefont {P.~V.}\ \bibnamefont
  {Kapitanova}}, \bibinfo {author} {\bibfnamefont {M.~V.}\ \bibnamefont
  {Rybin}}, \bibinfo {author} {\bibfnamefont {K.~L.}\ \bibnamefont {Koshelev}},
  \bibinfo {author} {\bibfnamefont {Z.~F.}\ \bibnamefont {Sadrieva}}, \bibinfo
  {author} {\bibfnamefont {K.~B.}\ \bibnamefont {Samusev}}, \bibinfo {author}
  {\bibfnamefont {Y.~S.}\ \bibnamefont {Kivshar}}, \ and\ \bibinfo {author}
  {\bibfnamefont {M.~F.}\ \bibnamefont {Limonov}},\ }\href {\doibase
  10.1088/1742-6596/1124/5/051058} {\bibfield  {journal} {\bibinfo  {journal}
  {Journal of Physics: Conference Series}\ }\textbf {\bibinfo {volume}
  {1124}},\ \bibinfo {pages} {51058} (\bibinfo {year} {2018})}\BibitemShut
  {NoStop}%
\bibitem [{\citenamefont {Khosravi}\ \emph {et~al.}(2019)\citenamefont
  {Khosravi}, \citenamefont {Cortes},\ and\ \citenamefont
  {Jacob}}]{KhosraviSO_2019}%
  \BibitemOpen
  \bibfield  {author} {\bibinfo {author} {\bibfnamefont {F.}~\bibnamefont
  {Khosravi}}, \bibinfo {author} {\bibfnamefont {C.~L.}\ \bibnamefont
  {Cortes}}, \ and\ \bibinfo {author} {\bibfnamefont {Z.}~\bibnamefont
  {Jacob}},\ }\href {\doibase 10.1364/OE.27.015846} {\bibfield  {journal}
  {\bibinfo  {journal} {Optics Express}\ }\textbf {\bibinfo {volume} {27}},\
  \bibinfo {pages} {15846} (\bibinfo {year} {2019})}\BibitemShut {NoStop}%
\bibitem [{\citenamefont {Bliokh}\ and\ \citenamefont
  {Nori}(2015)}]{BliokhSO_2015}%
  \BibitemOpen
  \bibfield  {author} {\bibinfo {author} {\bibfnamefont {K.~Y.}\ \bibnamefont
  {Bliokh}}\ and\ \bibinfo {author} {\bibfnamefont {F.}~\bibnamefont {Nori}},\
  }\href {\doibase 10.1016/j.physrep.2015.06.003} {\bibfield  {journal}
  {\bibinfo  {journal} {Physics Reports}\ }\textbf {\bibinfo {volume} {592}},\
  \bibinfo {pages} {1} (\bibinfo {year} {2015})}\BibitemShut {NoStop}%
\bibitem [{\citenamefont {Aiello}\ \emph {et~al.}(2015)\citenamefont {Aiello},
  \citenamefont {Banzer}, \citenamefont {Neugebauer},\ and\ \citenamefont
  {Leuchs}}]{AielloS0_2015}%
  \BibitemOpen
  \bibfield  {author} {\bibinfo {author} {\bibfnamefont {A.}~\bibnamefont
  {Aiello}}, \bibinfo {author} {\bibfnamefont {P.}~\bibnamefont {Banzer}},
  \bibinfo {author} {\bibfnamefont {M.}~\bibnamefont {Neugebauer}}, \ and\
  \bibinfo {author} {\bibfnamefont {G.}~\bibnamefont {Leuchs}},\ }\href
  {\doibase 10.1038/nphoton.2015.203} {\bibfield  {journal} {\bibinfo
  {journal} {Nature Photonics}\ }\textbf {\bibinfo {volume} {9}},\ \bibinfo
  {pages} {789} (\bibinfo {year} {2015})}\BibitemShut {NoStop}%
\bibitem [{\citenamefont {Cardano}\ and\ \citenamefont
  {Marrucci}(2015)}]{CardanoSO_2015}%
  \BibitemOpen
  \bibfield  {author} {\bibinfo {author} {\bibfnamefont {F.}~\bibnamefont
  {Cardano}}\ and\ \bibinfo {author} {\bibfnamefont {L.}~\bibnamefont
  {Marrucci}},\ }\href {http://dx.doi.org/10.1038/nphoton.2015.232} {\bibfield
  {journal} {\bibinfo  {journal} {Nat Photon}\ }\textbf {\bibinfo {volume}
  {9}},\ \bibinfo {pages} {776} (\bibinfo {year} {2015})}\BibitemShut {NoStop}%
\bibitem [{\citenamefont {Junge}\ \emph {et~al.}(2013)\citenamefont {Junge},
  \citenamefont {O'Shea}, \citenamefont {Volz},\ and\ \citenamefont
  {Rauschenbeutel}}]{JungeSO_2013}%
  \BibitemOpen
  \bibfield  {author} {\bibinfo {author} {\bibfnamefont {C.}~\bibnamefont
  {Junge}}, \bibinfo {author} {\bibfnamefont {D.}~\bibnamefont {O'Shea}},
  \bibinfo {author} {\bibfnamefont {J.}~\bibnamefont {Volz}}, \ and\ \bibinfo
  {author} {\bibfnamefont {A.}~\bibnamefont {Rauschenbeutel}},\ }\href
  {\doibase 10.1103/PhysRevLett.110.213604} {\bibfield  {journal} {\bibinfo
  {journal} {Physical Review Letters}\ }\textbf {\bibinfo {volume} {110}},\
  \bibinfo {pages} {1} (\bibinfo {year} {2013})},\ \Eprint
  {http://arxiv.org/abs/1301.1659} {arXiv:1301.1659} \BibitemShut {NoStop}%
\bibitem [{\citenamefont {O'Connor}\ \emph {et~al.}(2014)\citenamefont
  {O'Connor}, \citenamefont {Ginzburg}, \citenamefont
  {Rodr{\'{i}}guez-Fortu{\~{n}}o}, \citenamefont {Wurtz},\ and\ \citenamefont
  {Zayats}}]{OConnorSO_2014}%
  \BibitemOpen
  \bibfield  {author} {\bibinfo {author} {\bibfnamefont {D.}~\bibnamefont
  {O'Connor}}, \bibinfo {author} {\bibfnamefont {P.}~\bibnamefont {Ginzburg}},
  \bibinfo {author} {\bibfnamefont {F.~J.}\ \bibnamefont
  {Rodr{\'{i}}guez-Fortu{\~{n}}o}}, \bibinfo {author} {\bibfnamefont {G.~A.}\
  \bibnamefont {Wurtz}}, \ and\ \bibinfo {author} {\bibfnamefont {A.~V.}\
  \bibnamefont {Zayats}},\ }\href {\doibase 10.1038/ncomms6327} {\bibfield
  {journal} {\bibinfo  {journal} {Nature Communications}\ }\textbf {\bibinfo
  {volume} {5}},\ \bibinfo {pages} {1} (\bibinfo {year} {2014})}\BibitemShut
  {NoStop}%
\bibitem [{\citenamefont {Shao}\ \emph {et~al.}(2018)\citenamefont {Shao},
  \citenamefont {Zhu}, \citenamefont {Chen}, \citenamefont {Zhang},\ and\
  \citenamefont {Yu}}]{ShaoSO_2018}%
  \BibitemOpen
  \bibfield  {author} {\bibinfo {author} {\bibfnamefont {Z.}~\bibnamefont
  {Shao}}, \bibinfo {author} {\bibfnamefont {J.}~\bibnamefont {Zhu}}, \bibinfo
  {author} {\bibfnamefont {Y.}~\bibnamefont {Chen}}, \bibinfo {author}
  {\bibfnamefont {Y.}~\bibnamefont {Zhang}}, \ and\ \bibinfo {author}
  {\bibfnamefont {S.}~\bibnamefont {Yu}},\ }\href {\doibase
  10.1038/s41467-018-03237-5} {\bibfield  {journal} {\bibinfo  {journal}
  {Nature Communications}\ }\textbf {\bibinfo {volume} {9}},\ \bibinfo {pages}
  {1} (\bibinfo {year} {2018})},\ \Eprint {http://arxiv.org/abs/1709.09811}
  {arXiv:1709.09811} \BibitemShut {NoStop}%
\bibitem [{\citenamefont {Eismann}\ \emph {et~al.}(2019)\citenamefont
  {Eismann}, \citenamefont {Banzer},\ and\ \citenamefont
  {Neugebauer}}]{spinorbit-beams_2019}%
  \BibitemOpen
  \bibfield  {author} {\bibinfo {author} {\bibfnamefont {J.~S.}\ \bibnamefont
  {Eismann}}, \bibinfo {author} {\bibfnamefont {P.}~\bibnamefont {Banzer}}, \
  and\ \bibinfo {author} {\bibfnamefont {M.}~\bibnamefont {Neugebauer}},\
  }\href {\doibase 10.1103/physrevresearch.1.033143} {\bibfield  {journal}
  {\bibinfo  {journal} {Physical Review Research}\ }\textbf {\bibinfo {volume}
  {1}},\ \bibinfo {pages} {1} (\bibinfo {year} {2019})},\ \Eprint
  {http://arxiv.org/abs/1905.12539} {arXiv:1905.12539} \BibitemShut {NoStop}%
\bibitem [{\citenamefont {Rosenberger}\ \emph {et~al.}(2019)\citenamefont
  {Rosenberger}, \citenamefont {Dale}, \citenamefont {Bui}, \citenamefont
  {Gonzales}, \citenamefont {Ganta}, \citenamefont {Ke},\ and\ \citenamefont
  {Rajagopal}}]{RosenbergerSO_2019}%
  \BibitemOpen
  \bibfield  {author} {\bibinfo {author} {\bibfnamefont {A.~T.}\ \bibnamefont
  {Rosenberger}}, \bibinfo {author} {\bibfnamefont {E.~B.}\ \bibnamefont
  {Dale}}, \bibinfo {author} {\bibfnamefont {K.~V.}\ \bibnamefont {Bui}},
  \bibinfo {author} {\bibfnamefont {E.~K.}\ \bibnamefont {Gonzales}}, \bibinfo
  {author} {\bibfnamefont {D.}~\bibnamefont {Ganta}}, \bibinfo {author}
  {\bibfnamefont {L.}~\bibnamefont {Ke}}, \ and\ \bibinfo {author}
  {\bibfnamefont {S.~R.}\ \bibnamefont {Rajagopal}},\ }\href {\doibase
  10.1364/ol.44.004163} {\bibfield  {journal} {\bibinfo  {journal} {Optics
  Letters}\ }\textbf {\bibinfo {volume} {44}},\ \bibinfo {pages} {4163}
  (\bibinfo {year} {2019})}\BibitemShut {NoStop}%
\bibitem [{\citenamefont {Sun}\ \emph {et~al.}(2019)\citenamefont {Sun},
  \citenamefont {Bai}, \citenamefont {Wang}, \citenamefont {Zhang},
  \citenamefont {Zhang}, \citenamefont {Song},\ and\ \citenamefont
  {Huang}}]{SunSO_2019}%
  \BibitemOpen
  \bibfield  {author} {\bibinfo {author} {\bibfnamefont {L.}~\bibnamefont
  {Sun}}, \bibinfo {author} {\bibfnamefont {B.}~\bibnamefont {Bai}}, \bibinfo
  {author} {\bibfnamefont {J.}~\bibnamefont {Wang}}, \bibinfo {author}
  {\bibfnamefont {M.}~\bibnamefont {Zhang}}, \bibinfo {author} {\bibfnamefont
  {X.}~\bibnamefont {Zhang}}, \bibinfo {author} {\bibfnamefont
  {X.}~\bibnamefont {Song}}, \ and\ \bibinfo {author} {\bibfnamefont
  {L.}~\bibnamefont {Huang}},\ }\href {\doibase 10.1002/adfm.201902286}
  {\bibfield  {journal} {\bibinfo  {journal} {Advanced Functional Materials}\
  }\textbf {\bibinfo {volume} {29}},\ \bibinfo {pages} {1902286} (\bibinfo
  {year} {2019})}\BibitemShut {NoStop}%
\bibitem [{\citenamefont {Berry}(1984)}]{Berry1984}%
  \BibitemOpen
  \bibfield  {author} {\bibinfo {author} {\bibfnamefont {M.~V.}\ \bibnamefont
  {Berry}},\ }\href {\doibase 10.1098/rspa.1984.0023} {\bibfield  {journal}
  {\bibinfo  {journal} {Proceedings of the Royal Society of London A:
  Mathematical, Physical and Engineering Sciences}\ }\textbf {\bibinfo {volume}
  {392}},\ \bibinfo {pages} {45} (\bibinfo {year} {1984})}\BibitemShut
  {NoStop}%
\bibitem [{\citenamefont {Berry}(1987)}]{Berry1987}%
  \BibitemOpen
  \bibfield  {author} {\bibinfo {author} {\bibfnamefont {M.~V.}\ \bibnamefont
  {Berry}},\ }\href {\doibase 10.1038/326277a0} {\bibfield  {journal} {\bibinfo
   {journal} {Nature}\ }\textbf {\bibinfo {volume} {326}},\ \bibinfo {pages}
  {277} (\bibinfo {year} {1987})}\BibitemShut {NoStop}%
\bibitem [{\citenamefont {Bliokh}\ \emph {et~al.}(2008)\citenamefont {Bliokh},
  \citenamefont {Niv}, \citenamefont {Kleiner},\ and\ \citenamefont
  {Hasman}}]{Bliokh2008}%
  \BibitemOpen
  \bibfield  {author} {\bibinfo {author} {\bibfnamefont {K.~Y.}\ \bibnamefont
  {Bliokh}}, \bibinfo {author} {\bibfnamefont {A.}~\bibnamefont {Niv}},
  \bibinfo {author} {\bibfnamefont {V.}~\bibnamefont {Kleiner}}, \ and\
  \bibinfo {author} {\bibfnamefont {E.}~\bibnamefont {Hasman}},\ }\href
  {\doibase 10.1038/nphoton.2008.229} {\bibfield  {journal} {\bibinfo
  {journal} {Nature Photonics}\ }\textbf {\bibinfo {volume} {2}},\ \bibinfo
  {pages} {748} (\bibinfo {year} {2008})},\ \Eprint
  {http://arxiv.org/abs/0810.2136} {arXiv:0810.2136} \BibitemShut {NoStop}%
\bibitem [{\citenamefont {Bliokh}\ \emph {et~al.}(2019)\citenamefont {Bliokh},
  \citenamefont {Alonso},\ and\ \citenamefont {Dennis}}]{BliokhGP_2019}%
  \BibitemOpen
  \bibfield  {author} {\bibinfo {author} {\bibfnamefont {K.~Y.}\ \bibnamefont
  {Bliokh}}, \bibinfo {author} {\bibfnamefont {M.~A.}\ \bibnamefont {Alonso}},
  \ and\ \bibinfo {author} {\bibfnamefont {M.~R.}\ \bibnamefont {Dennis}},\
  }\href {\doibase 10.1088/1361-6633/ab4415} {\bibfield  {journal} {\bibinfo
  {journal} {Reports on progress in physics. Physical Society (Great Britain)}\
  }\textbf {\bibinfo {volume} {82}},\ \bibinfo {pages} {122401} (\bibinfo
  {year} {2019})},\ \Eprint {http://arxiv.org/abs/1903.01304}
  {arXiv:1903.01304} \BibitemShut {NoStop}%
\bibitem [{\citenamefont {Loss}\ \emph {et~al.}(1990)\citenamefont {Loss},
  \citenamefont {Goldbart},\ and\ \citenamefont {Balatsky}}]{Loss_1990}%
  \BibitemOpen
  \bibfield  {author} {\bibinfo {author} {\bibfnamefont {D.}~\bibnamefont
  {Loss}}, \bibinfo {author} {\bibfnamefont {P.}~\bibnamefont {Goldbart}}, \
  and\ \bibinfo {author} {\bibfnamefont {A.~V.}\ \bibnamefont {Balatsky}},\
  }\href {\doibase 10.1103/PhysRevLett.65.1655} {\bibfield  {journal} {\bibinfo
   {journal} {Physical Review Letters}\ }\textbf {\bibinfo {volume} {65}},\
  \bibinfo {pages} {1655} (\bibinfo {year} {1990})}\BibitemShut {NoStop}%
\bibitem [{\citenamefont {Nagasawa}\ \emph {et~al.}(2013)\citenamefont
  {Nagasawa}, \citenamefont {Frustaglia}, \citenamefont {Saarikoski},
  \citenamefont {Richter},\ and\ \citenamefont {Nitta}}]{Nagasawa_2013}%
  \BibitemOpen
  \bibfield  {author} {\bibinfo {author} {\bibfnamefont {F.}~\bibnamefont
  {Nagasawa}}, \bibinfo {author} {\bibfnamefont {D.}~\bibnamefont
  {Frustaglia}}, \bibinfo {author} {\bibfnamefont {H.}~\bibnamefont
  {Saarikoski}}, \bibinfo {author} {\bibfnamefont {K.}~\bibnamefont {Richter}},
  \ and\ \bibinfo {author} {\bibfnamefont {J.}~\bibnamefont {Nitta}},\ }\href
  {\doibase 10.1038/ncomms3526} {\bibfield  {journal} {\bibinfo  {journal}
  {Nature Communications}\ }\textbf {\bibinfo {volume} {4}} (\bibinfo {year}
  {2013}),\ 10.1038/ncomms3526}\BibitemShut {NoStop}%
\bibitem [{\citenamefont {Frustaglia}\ \emph {et~al.}(2001)\citenamefont
  {Frustaglia}, \citenamefont {Hentschel},\ and\ \citenamefont
  {Richter}}]{Frustaglia2001}%
  \BibitemOpen
  \bibfield  {author} {\bibinfo {author} {\bibfnamefont {D.}~\bibnamefont
  {Frustaglia}}, \bibinfo {author} {\bibfnamefont {M.}~\bibnamefont
  {Hentschel}}, \ and\ \bibinfo {author} {\bibfnamefont {K.}~\bibnamefont
  {Richter}},\ }\href {\doibase 10.1103/PhysRevLett.87.256602} {\bibfield
  {journal} {\bibinfo  {journal} {Phys. Rev. Lett.}\ }\textbf {\bibinfo
  {volume} {87}},\ \bibinfo {pages} {256602} (\bibinfo {year}
  {2001})}\BibitemShut {NoStop}%
\bibitem [{\citenamefont {Hentschel}\ \emph {et~al.}(2004)\citenamefont
  {Hentschel}, \citenamefont {Schomerus}, \citenamefont {Frustaglia},\ and\
  \citenamefont {Richter}}]{Hentschel2004}%
  \BibitemOpen
  \bibfield  {author} {\bibinfo {author} {\bibfnamefont {M.}~\bibnamefont
  {Hentschel}}, \bibinfo {author} {\bibfnamefont {H.}~\bibnamefont
  {Schomerus}}, \bibinfo {author} {\bibfnamefont {D.}~\bibnamefont
  {Frustaglia}}, \ and\ \bibinfo {author} {\bibfnamefont {K.}~\bibnamefont
  {Richter}},\ }\href {\doibase 10.1103/PhysRevB.69.155326} {\bibfield
  {journal} {\bibinfo  {journal} {Phys. Rev. B}\ }\textbf {\bibinfo {volume}
  {69}},\ \bibinfo {pages} {155326} (\bibinfo {year} {2004})}\BibitemShut
  {NoStop}%
\bibitem [{\citenamefont {Jackson}(1998)}]{Jackson}%
  \BibitemOpen
  \bibinfo {editor} {\bibfnamefont {J.~D.}\ \bibnamefont {Jackson}},\ ed.,\
  \href@noop {} {\emph {\bibinfo {title} {{Classical Electrodynamics}}}},\
  \bibinfo {edition} {3rd}\ ed.\ (\bibinfo  {publisher} {Wiley, New York},\
  \bibinfo {year} {1998})\BibitemShut {NoStop}%
\bibitem [{\citenamefont {{Bola{\~{n}}os Qui{\~{n}}ones}}\ \emph
  {et~al.}(2009)\citenamefont {{Bola{\~{n}}os Qui{\~{n}}ones}}, \citenamefont
  {Huang}, \citenamefont {Plumhof}, \citenamefont {Kiravittaya}, \citenamefont
  {Rastelli}, \citenamefont {Mei},\ and\ \citenamefont
  {Schmidt}}]{wgm-tubes_2009}%
  \BibitemOpen
  \bibfield  {author} {\bibinfo {author} {\bibfnamefont {V.~A.}\ \bibnamefont
  {{Bola{\~{n}}os Qui{\~{n}}ones}}}, \bibinfo {author} {\bibfnamefont
  {G.}~\bibnamefont {Huang}}, \bibinfo {author} {\bibfnamefont {J.~D.}\
  \bibnamefont {Plumhof}}, \bibinfo {author} {\bibfnamefont {S.}~\bibnamefont
  {Kiravittaya}}, \bibinfo {author} {\bibfnamefont {A.}~\bibnamefont
  {Rastelli}}, \bibinfo {author} {\bibfnamefont {Y.}~\bibnamefont {Mei}}, \
  and\ \bibinfo {author} {\bibfnamefont {O.~G.}\ \bibnamefont {Schmidt}},\
  }\href {\doibase 10.1364/ol.34.002345} {\bibfield  {journal} {\bibinfo
  {journal} {Optics Letters}\ }\textbf {\bibinfo {volume} {34}},\ \bibinfo
  {pages} {2345} (\bibinfo {year} {2009})}\BibitemShut {NoStop}%
\bibitem [{\citenamefont {{Bola{\~{n}}os Qui{\~{n}}ones}}\ \emph
  {et~al.}(2012)\citenamefont {{Bola{\~{n}}os Qui{\~{n}}ones}}, \citenamefont
  {Ma}, \citenamefont {Li}, \citenamefont {Jorgensen}, \citenamefont
  {Kiravittaya},\ and\ \citenamefont {Schmidt}}]{Belanos_2012}%
  \BibitemOpen
  \bibfield  {author} {\bibinfo {author} {\bibfnamefont {V.~A.}\ \bibnamefont
  {{Bola{\~{n}}os Qui{\~{n}}ones}}}, \bibinfo {author} {\bibfnamefont
  {L.}~\bibnamefont {Ma}}, \bibinfo {author} {\bibfnamefont {S.}~\bibnamefont
  {Li}}, \bibinfo {author} {\bibfnamefont {M.}~\bibnamefont {Jorgensen}},
  \bibinfo {author} {\bibfnamefont {S.}~\bibnamefont {Kiravittaya}}, \ and\
  \bibinfo {author} {\bibfnamefont {O.~G.}\ \bibnamefont {Schmidt}},\ }\href
  {\doibase 10.1364/OL.37.004284} {\bibfield  {journal} {\bibinfo  {journal}
  {Optics Letters}\ }\textbf {\bibinfo {volume} {37}},\ \bibinfo {pages} {4284}
  (\bibinfo {year} {2012})}\BibitemShut {NoStop}%
\bibitem [{\citenamefont {Collett}(2005)}]{polarization_ellipse}%
  \BibitemOpen
  \bibfield  {author} {\bibinfo {author} {\bibfnamefont {E.}~\bibnamefont
  {Collett}},\ }\href@noop {} {\emph {\bibinfo {title} {{Field Guide to
  Polarization}}}}\ (\bibinfo  {publisher} {SPIE Press, Bellingham},\ \bibinfo
  {year} {2005})\BibitemShut {NoStop}%
\bibitem [{\citenamefont {Oskooi}\ \emph {et~al.}(2010)\citenamefont {Oskooi},
  \citenamefont {Roundy}, \citenamefont {Ibanescu}, \citenamefont {Bermel},
  \citenamefont {Joannopoulos},\ and\ \citenamefont {Johnson}}]{MEEP}%
  \BibitemOpen
  \bibfield  {author} {\bibinfo {author} {\bibfnamefont {A.~F.}\ \bibnamefont
  {Oskooi}}, \bibinfo {author} {\bibfnamefont {D.}~\bibnamefont {Roundy}},
  \bibinfo {author} {\bibfnamefont {M.}~\bibnamefont {Ibanescu}}, \bibinfo
  {author} {\bibfnamefont {P.}~\bibnamefont {Bermel}}, \bibinfo {author}
  {\bibfnamefont {J.~D.}\ \bibnamefont {Joannopoulos}}, \ and\ \bibinfo
  {author} {\bibfnamefont {S.~G.}\ \bibnamefont {Johnson}},\ }\href@noop {}
  {\bibfield  {journal} {\bibinfo  {journal} {Computer Physics Communications}\
  }\textbf {\bibinfo {volume} {181}},\ \bibinfo {pages} {687} (\bibinfo {year}
  {2010})}\BibitemShut {NoStop}%
\end{thebibliography}%


\begin{thebibliography}{7}%
\makeatletter
\providecommand \@ifxundefined [1]{%
 \@ifx{#1\undefined}
}%
\providecommand \@ifnum [1]{%
 \ifnum #1\expandafter \@firstoftwo
 \else \expandafter \@secondoftwo
 \fi
}%
\providecommand \@ifx [1]{%
 \ifx #1\expandafter \@firstoftwo
 \else \expandafter \@secondoftwo
 \fi
}%
\providecommand \natexlab [1]{#1}%
\providecommand \enquote  [1]{``#1''}%
\providecommand \bibnamefont  [1]{#1}%
\providecommand \bibfnamefont [1]{#1}%
\providecommand \citenamefont [1]{#1}%
\providecommand \href@noop [0]{\@secondoftwo}%
\providecommand \href [0]{\begingroup \@sanitize@url \@href}%
\providecommand \@href[1]{\@@startlink{#1}\@@href}%
\providecommand \@@href[1]{\endgroup#1\@@endlink}%
\providecommand \@sanitize@url [0]{\catcode `\\12\catcode `\$12\catcode
  `\&12\catcode `\#12\catcode `\^12\catcode `\_12\catcode `\%12\relax}%
\providecommand \@@startlink[1]{}%
\providecommand \@@endlink[0]{}%
\providecommand \url  [0]{\begingroup\@sanitize@url \@url }%
\providecommand \@url [1]{\endgroup\@href {#1}{\urlprefix }}%
\providecommand \urlprefix  [0]{URL }%
\providecommand \Eprint [0]{\href }%
\providecommand \doibase [0]{http://dx.doi.org/}%
\providecommand \selectlanguage [0]{\@gobble}%
\providecommand \bibinfo  [0]{\@secondoftwo}%
\providecommand \bibfield  [0]{\@secondoftwo}%
\providecommand \translation [1]{[#1]}%
\providecommand \BibitemOpen [0]{}%
\providecommand \bibitemStop [0]{}%
\providecommand \bibitemNoStop [0]{.\EOS\space}%
\providecommand \EOS [0]{\spacefactor3000\relax}%
\providecommand \BibitemShut  [1]{\csname bibitem#1\endcsname}%
\let\auto@bib@innerbib\@empty
\bibitem [{\citenamefont {Jackson}(1998)}]{Jackson}%
  \BibitemOpen
  \bibinfo {editor} {\bibfnamefont {J.~D.}\ \bibnamefont {Jackson}},\ ed.,\
  \href@noop {} {\emph {\bibinfo {title} {{Classical Electrodynamics}}}},\
  \bibinfo {edition} {3rd}\ ed.\ (\bibinfo  {publisher} {Wiley, New York},\
  \bibinfo {year} {1998})\BibitemShut {NoStop}%
\bibitem [{\citenamefont {{David Colton}}(1998)}]{JacobiAnger}%
  \BibitemOpen
  \bibfield  {author} {\bibinfo {author} {\bibfnamefont {R.~K.}\ \bibnamefont
  {{David Colton}}},\ }\href@noop {} {\emph {\bibinfo {title} {{Inverse
  Acoustic and Electromagnetic Scattering Theory}}}}\ (\bibinfo  {publisher}
  {Springer, Berlin, Heidelberg},\ \bibinfo {year} {1998})\ p.~\bibinfo {pages}
  {32}\BibitemShut {NoStop}%
\bibitem [{\citenamefont {{Alfred Shapere}}(1989)}]{geomphaseBook_1989}%
  \BibitemOpen
  \bibfield  {author} {\bibinfo {author} {\bibfnamefont {F.~W.}\ \bibnamefont
  {{Alfred Shapere}}},\ }\href@noop {} {\emph {\bibinfo {title} {{GEOMETRIC
  PHASES IN PHYSICS Vol. 5}}}}\ (\bibinfo  {publisher} {World Scientific,
  Singapore},\ \bibinfo {year} {1989})\ pp.\ \bibinfo {pages} {45--104,
  193--240}\BibitemShut {NoStop}%
\bibitem [{\citenamefont {Loss}\ \emph {et~al.}(1990)\citenamefont {Loss},
  \citenamefont {Goldbart},\ and\ \citenamefont {Balatsky}}]{Loss_1990}%
  \BibitemOpen
  \bibfield  {author} {\bibinfo {author} {\bibfnamefont {D.}~\bibnamefont
  {Loss}}, \bibinfo {author} {\bibfnamefont {P.}~\bibnamefont {Goldbart}}, \
  and\ \bibinfo {author} {\bibfnamefont {A.~V.}\ \bibnamefont {Balatsky}},\
  }\href {\doibase 10.1103/PhysRevLett.65.1655} {\bibfield  {journal} {\bibinfo
   {journal} {Physical Review Letters}\ }\textbf {\bibinfo {volume} {65}},\
  \bibinfo {pages} {1655} (\bibinfo {year} {1990})}\BibitemShut {NoStop}%
\bibitem [{\citenamefont {Nagasawa}\ \emph {et~al.}(2013)\citenamefont
  {Nagasawa}, \citenamefont {Frustaglia}, \citenamefont {Saarikoski},
  \citenamefont {Richter},\ and\ \citenamefont {Nitta}}]{Nagasawa_2013}%
  \BibitemOpen
  \bibfield  {author} {\bibinfo {author} {\bibfnamefont {F.}~\bibnamefont
  {Nagasawa}}, \bibinfo {author} {\bibfnamefont {D.}~\bibnamefont
  {Frustaglia}}, \bibinfo {author} {\bibfnamefont {H.}~\bibnamefont
  {Saarikoski}}, \bibinfo {author} {\bibfnamefont {K.}~\bibnamefont {Richter}},
  \ and\ \bibinfo {author} {\bibfnamefont {J.}~\bibnamefont {Nitta}},\ }\href
  {\doibase 10.1038/ncomms3526} {\bibfield  {journal} {\bibinfo  {journal}
  {Nature Communications}\ }\textbf {\bibinfo {volume} {4}} (\bibinfo {year}
  {2013}),\ 10.1038/ncomms3526}\BibitemShut {NoStop}%
\bibitem [{\citenamefont {Frustaglia}\ \emph {et~al.}(2001)\citenamefont
  {Frustaglia}, \citenamefont {Hentschel},\ and\ \citenamefont
  {Richter}}]{Frustaglia2001}%
  \BibitemOpen
  \bibfield  {author} {\bibinfo {author} {\bibfnamefont {D.}~\bibnamefont
  {Frustaglia}}, \bibinfo {author} {\bibfnamefont {M.}~\bibnamefont
  {Hentschel}}, \ and\ \bibinfo {author} {\bibfnamefont {K.}~\bibnamefont
  {Richter}},\ }\href {\doibase 10.1103/PhysRevLett.87.256602} {\bibfield
  {journal} {\bibinfo  {journal} {Phys. Rev. Lett.}\ }\textbf {\bibinfo
  {volume} {87}},\ \bibinfo {pages} {256602} (\bibinfo {year}
  {2001})}\BibitemShut {NoStop}%
\bibitem [{\citenamefont {Hentschel}\ \emph {et~al.}(2004)\citenamefont
  {Hentschel}, \citenamefont {Schomerus}, \citenamefont {Frustaglia},\ and\
  \citenamefont {Richter}}]{Hentschel2004}%
  \BibitemOpen
  \bibfield  {author} {\bibinfo {author} {\bibfnamefont {M.}~\bibnamefont
  {Hentschel}}, \bibinfo {author} {\bibfnamefont {H.}~\bibnamefont
  {Schomerus}}, \bibinfo {author} {\bibfnamefont {D.}~\bibnamefont
  {Frustaglia}}, \ and\ \bibinfo {author} {\bibfnamefont {K.}~\bibnamefont
  {Richter}},\ }\href {\doibase 10.1103/PhysRevB.69.155326} {\bibfield
  {journal} {\bibinfo  {journal} {Phys. Rev. B}\ }\textbf {\bibinfo {volume}
  {69}},\ \bibinfo {pages} {155326} (\bibinfo {year} {2004})}\BibitemShut
  {NoStop}%
\end{thebibliography}%

\end{document}


\begin{center}
{\Large{\textbf{Supplemental Material }}} \\ 
for \textquotedblleft Spin-Orbit interaction of light in 3D microcavities\textquotedblright	
\end{center}

\section{Vector diffraction theory}
\label{sec:1}

We start by recalling the definition of the electric far-field field vector according to Kirchhoff vector diffraction~\cite{Jackson}:
\begin{equation}
\textbf{E}_\text{FF}(\varphi,\theta)\propto \textbf{k}(\varphi,\theta)\times\iint \textbf{n}(\textbf{x}')\times\textbf{E}(\textbf{x}')e^{-i \textbf{k}\cdot\textbf{x}'}\d a' = \textbf{k}(\varphi,\theta)\times\textbf{K}(\varphi,\theta)
\end{equation}
\begin{figure}[h!]
	\centering
	\includegraphics[width=12cm]{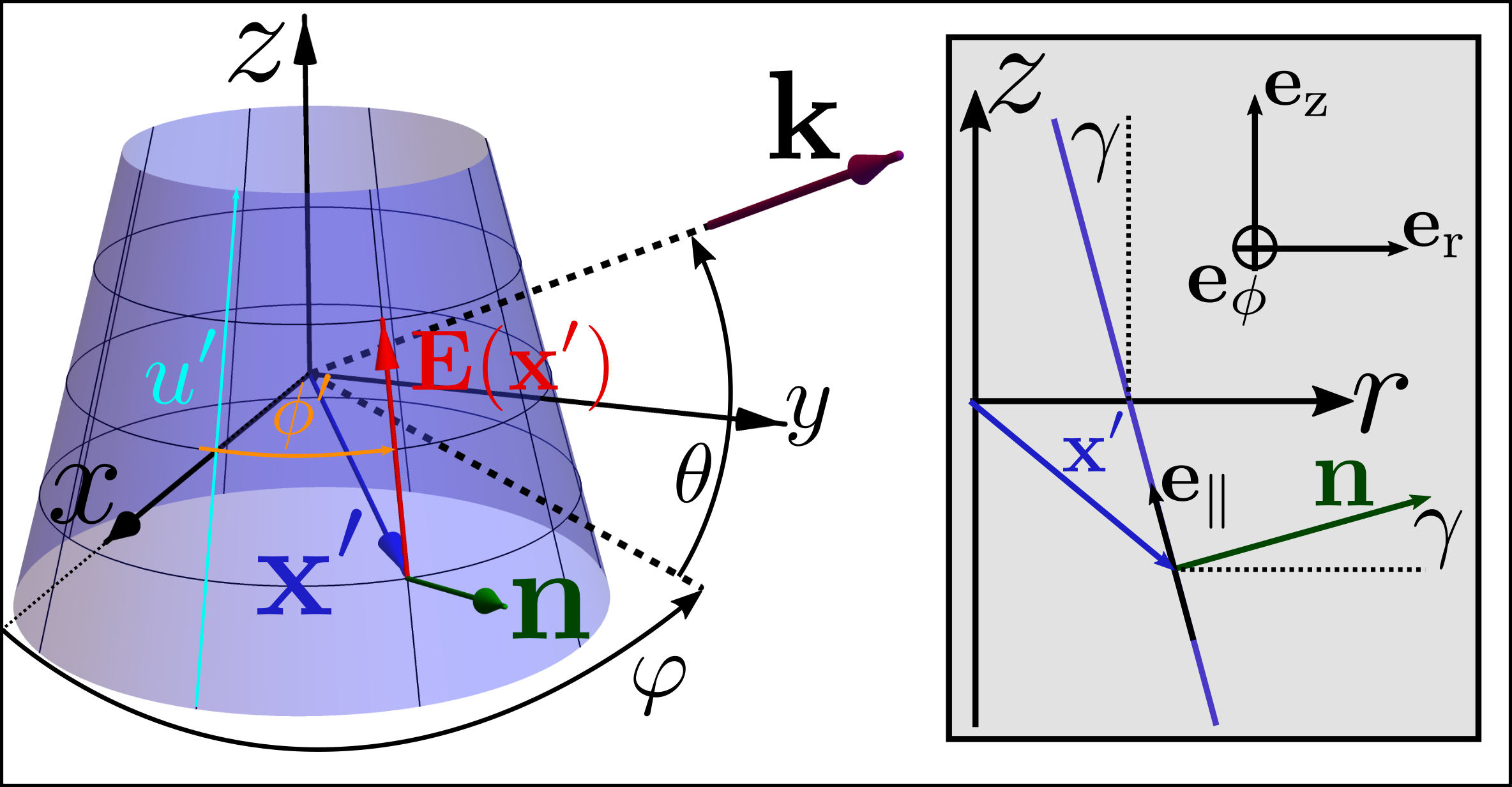}
	\caption{\textbf{Illustration of the geometry:} The blue cone surface is parameterized by ($u'$, $\phi'$). $(\theta,\varphi)$ are the far-field angles. $\gamma$ is the angle between the cone surface and the $z$-axis.}
	\label{fig:S1}
\end{figure}
with  $\textbf{x}'=\textbf{x}'(u',\phi')$ being the position vector pointing onto the surface $S$ parameterized by the elevation parameter $u'$ and the azimuthal parameter $\phi'$, cf.~Fig.~\ref{fig:S1}. The vector $\textbf{n}$ is the unit normal vector of the surface $S$, $\d a'$ is the differential surface area element given by $\big|\frac{\partial\textbf{x}'}{\partial u'}\times\frac{\partial\textbf{x}'}{\partial \phi'}\big|\d u'\d \phi'=s(u',\phi')\d u'\d \phi' $ and $\textbf{k}(\varphi,\theta)$ represents the wave vector pointing into the observation direction $(\varphi, \theta)$ in the far field, $\textbf{k} = k \left(\cos\theta\cos\varphi~\textbf{e}_\text{x} + \cos\theta\sin\varphi~\textbf{e}_\text{y} +\sin\theta~\textbf{e}_\text{z}   \right)$ , cf.~Fig.~1 of the manuscript. Thus, the scalar product $\textbf{k}\cdot\textbf{x}'=g(u',\phi',\theta,\varphi)$ is a function of $\theta$, $\varphi$, $u'$ and $\phi'$.\\
\subsection{Vector diffraction theory for cylindrical ring cavities ($\gamma=0$) }
\label{sec:diff_ring}
\noindent First, we investigate the simple case of a cylindrical ring cavity with $\gamma=0$ that is the wall of the ring cavity is aligned parallel to the $z$-axis, cf. Fig.~2 \textbf{a} of the manuscript.\\
The position vector can be written as $\textbf{x}'_\text{ring}(u',\phi')=R\textbf{e}_\text{r}(\phi') + h u'\textbf{e}_\text{z}$ where $\textbf{e}_\text{r}$ (pointing away radially from the $z$-axis) and $\textbf{e}_\text{z}$ (pointing parallel to the $z$-axis) are  unit vectors of cylindrical coordinates, cf. inset of Fig.~\ref{fig:S1}. The parameter $u'$ ranges from $-1/2$ to $1/2$ and $\phi'$ ranges from $0$ to $2\pi$. The normal vector of $S$ coincides with $\textbf{e}_\text{r}$, $\textbf{n}=\textbf{e}_\text{r}$. The electric fields at the cylindrical diffracting surface can be represented as $\textbf{E}(\textbf{x}_\text{ring}') = \left(A_\text{z}(z')\textbf{e}_\text{z} + A_\phi(z')\textbf{e}_\phi \right)\exp{(im\phi')} = \left(\tilde{A}_\text{z}(u')\textbf{e}_\text{z} + \tilde{A}_\phi(u')\textbf{e}_\phi \right)\exp{(im\phi')}  $. Thus, the term $\textbf{n}(\textbf{x}'_\text{ring})\times\textbf{E}(\textbf{x}'_\text{ring})$ becomes:
\begin{equation}\label{equ:nxE_ring}
	\textbf{n}(\textbf{x}'_\text{ring})\times\textbf{E}(\textbf{x}'_\text{ring})=\textbf{e}_\text{r}(\phi')\times\textbf{E}(u',\phi') = \left(\tilde{A}_\phi(u')\textbf{e}_\text{z} - \tilde{A}_\text{z}(u')\textbf{e}_\phi(\phi') \right)\exp{\left(im\phi'\right)}
\end{equation}
The scalar product $\textbf{k}\cdot\textbf{x}'_\text{ring}$ separates into two terms depending on $\phi'$ and $u'$, respectively:
\begin{equation} \label{equ:kx_ring}
\textbf{k}\cdot\textbf{x}'_\text{ring}= kR\cos\theta\cos\left(\phi'-\varphi\right) + kh\sin\theta~u'= \tilde{k}_1\cos\left(\phi'-\phi\right)+\tilde{k}_2u'.
\end{equation}
In the following, we will evaluate the components $K_\text{x}$, $K_\text{y}$ and $K_\text{z}$  of the diffraction integral $\textbf{K}$:
\begin{align}
K_\text{x}	&= h R \int_{0}^{2\pi}\sin\phi'\exp{(i m\phi')}\exp{(-i\tilde{k}_1\cos(\phi'-\varphi))}\text{d}\phi'\cdot \int_{-1/2}^{1/2}\tilde{A}_\text{z}(u')\exp{(-i\tilde{k}_2 u')}\text{d}u' \\
&= h R~\hat{\Phi}_\text{s}\left(\tilde{k}_1,\varphi\right) \cdot \hat{A}_\text{z}\left(\tilde{k}_2\right)	\\
K_\text{y}	&= - h R \int_{0}^{2\pi}\cos\phi'\exp{(i m\phi')}\exp{(-i\tilde{k}_1\cos(\phi'-\varphi))}\text{d}\phi'\cdot \int_{-1/2}^{1/2}\tilde{A}_\text{z}(u')\exp{(-i\tilde{k}_2 u')}\text{d}u' \\
&= -h R~\hat{\Phi}_\text{c}\left(\tilde{k}_1,\varphi\right) \cdot \hat{A}_\text{z}\left(\tilde{k}_2\right)	\\
K_\text{z} &= h R \int_{0}^{2\pi}\exp{(i m\phi')}\exp{(-i\tilde{k}_1\cos(\phi'-\varphi))}\text{d}\phi'\cdot \int_{-1/2}^{1/2}\tilde{A}_\phi(u')\exp{(-i\tilde{k}_2 u')}\text{d}u' \\[5pt]
&= i \frac{R^2}{m} \int_{0}^{2\pi}\exp{(i m\phi')}\exp{(-i\tilde{k}_1\cos(\phi'-\varphi))}\text{d}\phi'\cdot \int_{-1/2}^{1/2}\frac{\partial\tilde{A}_\text{z}(u')}{\partial u'}\exp{(-i\tilde{k}_2 u')}\text{d}u' \\[5pt]
&= \frac{R^2}{m} \hat{\Phi}_\text{1}\left(\tilde{k}_1,\varphi\right)\cdot \bigg(-\tilde{k}_2\hat{A}_\text{z}\left(\tilde{k}_2\right) + i \tilde{A}_\text{z}(u')\exp{\left(-i\tilde{k}_2 u'\right)}\bigg|_{u_1'=-1/2}^{u_2'=1/2} \bigg)
\end{align}
where we used $A_\phi(z')=i (R/m) (\partial A_\text{z}(z')/\partial z')$ and, therefore, $\tilde{A}_\phi(u')=i(R/mh) (\partial\tilde{A}_\text{z}(u')/\partial u')$ and integrated by parts.\\
In the next step, we simplify the $\hat{\Phi}_\text{s}\left(\tilde{k}_1,\varphi\right)$, $\hat{\Phi}_\text{c}\left(\tilde{k}_1,\varphi\right)$ and $\hat{\Phi}_1 \left(\tilde{k}_1,\varphi\right)$ using the substitution $\phi'=\phi+\varphi$ and taking advantage of angle sum identities of sine and cosine:
\begin{align}
\hat{\Phi}_\text{s}\left(\tilde{k}_1,\varphi\right)		&= \exp{(im\varphi)}\bigg(\cos\varphi~I_\text{s} + \sin\varphi~I_\text{c}  \bigg) \\
\hat{\Phi}_\text{c}\left(\tilde{k}_1,\varphi\right)		&= \exp{(im\varphi)}\bigg(\cos\varphi~I_\text{c} - \sin\varphi~I_\text{s}  \bigg) \\
\hat{\Phi}_1\left(\tilde{k}_1,\varphi\right)		&= \exp{(im\varphi)}I_1 
\end{align}
where the $I_\text{s}$, $I_\text{c}$ and $I_1$ are given by:
\begin{align}
I_\text{s}		&= \int_{0}^{2\pi}\sin\phi\exp{(im\phi)}\exp{\left(-i\tilde{k}_1\cos\phi\right)}\text{d}\phi = - \exp{\left(-i\frac{\pi}{2}m\right)} \frac{m}{\tilde{k}_1}J_m\left(\tilde{k}_1\right) \\
I_\text{c}		&= 	\int_{0}^{2\pi}\cos\phi\exp{(im\phi)}\exp{\left(-i\tilde{k}_1\cos\phi\right)}\text{d}\phi = \exp{\left(-i\frac{\pi}{2}(m+1)\right)}\frac{J_{m+1}\left(\tilde{k}_1\right)-J_{m-1}\left(\tilde{k}_1\right)}{2} \\	
I_1		&= \int_{0}^{2\pi}\exp{(im\phi)}\exp{\left(-i\tilde{k}_1\cos\phi\right)}\text{d}\phi = \exp{\left(-i\frac{\pi}{2}m\right)} J_m\left(\tilde{k}_1\right). 
\end{align}
The integrals $I_\text{s}$, $I_\text{c}$ and $I_1$ were evaluated using the Jacobi-Anger expansion~\cite{JacobiAnger}:
\begin{equation}
\exp{(i z \cos\phi)}=\sum_{n=-\infty}^{+\infty} \exp{\left(i\frac{\pi}{2}n\right)}J_n(z)\exp{(in\phi)}
\end{equation}
where the $J_n$ are the Bessel functions of the first kind.\\
Now, we derive the far-field components $E_\varphi$ and $E_\theta$:
\begin{align}
E_\varphi		&= \textbf{E}_\text{FF}\cdot\textbf{e}_\varphi= \left(\textbf{k}\times\textbf{K}\right)\cdot\textbf{e}_\varphi = -\textbf{K}\cdot\textbf{e}_\theta = \left(K_\text{x}\cos\varphi + K_\text{y}\sin\varphi\right)\sin\theta - K_\text{z}\cos\theta \nonumber\\
&= h R \exp{(im\varphi)}\exp{\left(-i\frac{\pi}{2}m\right)}J_m\left(\tilde{k}_1\right)\biggl(\hat{A}_\text{z}\left(\tilde{k}_2\right)\frac{\tilde{k}_1^2-m^2}{m\tilde{k}_1}\sin\theta - \Delta\cos\theta \biggr) \\
E_\theta		&= \textbf{E}_\text{FF}\cdot\textbf{e}_\theta= \left(\textbf{k}\times\textbf{K}\right)\cdot\textbf{e}_\theta = \textbf{K}\cdot\textbf{e}_\varphi = - K_\text{x}\sin\varphi + K_\text{y}\cos\varphi \nonumber\\
&= h R \exp{(im\varphi)\exp{\left(-i\frac{\pi}{2}(m+1)\right)}}\frac{J_{m-1}\left(\tilde{k}_1\right) - J_{m+1}\left(\tilde{k}_1\right) }{2}\hat{A}_\text{z}\left(\tilde{k}_2\right)
\end{align}
where $\Delta$ represents the boundary contributions 
\begin{align}
\Delta  ={} &  \frac{i}{m}\frac{R}{h}\tilde{A}_\text{z}(u')\exp{\left(-i\tilde{k}_2 u'\right)}\bigg|_{u_1'=-1/2}^{u_2'=1/2}  \\[5pt] 
={} & \frac{2R}{m h}\tilde{A}_\text{z}(1/2) \cdot \begin{cases}
\sin\left(\tilde{k}_2/2\right) &\text{for $q=1,3,5,...$}\\[10pt]
i\cos\left(\tilde{k}_2/2\right) &\text{for $q=2,4,6,...$}
\end{cases}
\end{align}

\newpage
\subsection{Vector diffraction theory for cone-like ring and tubular cavities ($\gamma>0$) }
\noindent In this subsection, we investigate the more complicated situation of an inclined cavity wall that is $\gamma > 0$.\\ 
The normal vector $\textbf{n}$ of the surface $S$ representing a thin-walled conical cavity can be written as 
\begin{equation*}
\textbf{n} = \cos\gamma~\textbf{e}_\text{r} + \sin\gamma~\textbf{e}_\text{z} 
\label{eq_n}
\end{equation*}
where $\gamma$ is the angle between the cone's surface and the $z$-axis or the half opening angle of the cone. Thus, the term $\textbf{n}\times\textbf{E}$ becomes
\begin{align*}
\textbf{n}\times\textbf{E} &={} \cos\gamma~\textbf{e}_\text{r}\times\textbf{E} +  \sin\gamma~\textbf{e}_\text{z}\times\textbf{E} \\
&={} \cos\gamma~\textbf{e}_\text{r}\times\textbf{E} + \sin\gamma~\bigg(-i m \textbf{E} + \frac{\partial \textbf{E}}{\partial \phi'} \bigg) \:.
\end{align*}
Here, we took advantage of the fact that the WGM is propagating on a circular trajectory around the cone (or $z$) axis. 
As a consequence, the triad of its moving frame is precessing around the $z$-axis. The direction vectors of the triad of the moving frame coincide (locally) with the direction vectors of cylindrical coordinates.
The change of the electric field vector along its path around the axis of the cone can be written as:
\begin{equation*}
\frac{\partial\textbf{E}}{\partial\phi'}= i m \textbf{E} + \sum_{j=r,\phi,z} E_j \frac{\partial\textbf{e}_j}{\partial\phi'} = i m \textbf{E} + \textbf{e}_\text{z}\times\textbf{E} 
\end{equation*}
where the change of the unit direction vectors $\frac{\partial\textbf{e}_j}{\partial\phi'}$ represents the precession of the triad around the $z$-axis, and therefore is equal to $\textbf{e}_\text{z}\times\textbf{E} $.\\
Inserting this into the definition of the electric far-field vector, we obtain
\begin{equation*}
\textbf{E}_\text{FF}=\textbf{k}\times\bigg[\cos\gamma\iint \textbf{e}_\text{r}\times\textbf{E}~e^{-i\textbf{k}\cdot\textbf{x}'}\d a' + \sin\gamma\iint\bigg(-i m \textbf{E} + \frac{\partial\textbf{E}}{\partial\phi'}  \bigg)~e^{-i\textbf{k}\cdot\textbf{x}'}\d a'  \bigg].
\end{equation*}
The integration over the cone area $\d a'=s(u',\phi')\d u'\d \phi'$ can be interpreted to be comprised of (i) an integration along the trajectory of the WGM represented by the  $\int (...)  \d\phi'$-integral and (ii) an integration perpendicular to the trajectory (along the height of the cavity) represented by the $\int (...)  \d u'$-integral. 

In the second term on the right-hand side we can integrate by parts the $\int (...)  \d\phi'$-integral, 
\begin{multline*}
\iint\frac{\partial\textbf{E}}{\partial\phi'}~e^{-i\textbf{k}\cdot\textbf{x}'}\d a'  = \\  \int\textbf{E}(u',\phi')~e^{-i\textbf{k}\cdot\textbf{x}'}s(u',\phi')\bigg|_{\phi'=0}^{\phi'=2\pi}~\d u' - \iint \bigg(-i\textbf{k}\cdot \frac{\partial\textbf{x}'}{\partial\phi'} + \frac{1}{s}\frac{\partial s}{\partial \phi'} \bigg)\textbf{E}~e^{-i\textbf{k}\cdot\textbf{x}'}\d a'.
\end{multline*}
Note that the last term inside the integral on the right-hand side vanishes for cylindrical symmetry, $ \partial s /\partial \phi' = 0$. 
The first term  depends on the  electric field of the start ($\phi'=0$) and end  ($\phi'=2\pi$) point of the trajectory around the cone axis. In the case of a WGM as a cyclical, stationary phenomenon, this difference vanishes because the electric field at the start and the end is the same or phase shifted by integer multiples of $2\pi$ (constructive interference condition). However, this term may contribute in more complex, non-cyclical wave dynamics.\\
\noindent Finally, the electric far-field vector can be written as
\begin{align}
& \textbf{E}_\text{FF}(\theta,\varphi) ={} \textbf{k}(\theta,\varphi)\times\textbf{K}(\varphi,\theta) =  \nonumber \\ 
& \textbf{k}(\theta,\varphi)\times\bigg[\cos\gamma\iint \textbf{e}_\text{r}\times\textbf{E}(u',\phi')~e^{-i\textbf{k}\cdot\textbf{x}'}\d a'~-~i\sin\gamma\iint \bigg(m-\textbf{k}\cdot\frac{\partial\textbf{x}'}{\partial\phi'} \bigg)\textbf{E}(u',\phi')~e^{-i\textbf{k}\cdot\textbf{x}'}\d a' \bigg] \\
& \quad\quad\quad\quad ={}  \textbf{k}(\theta,\varphi)\times\bigg(\cos\gamma~\textbf{K}_\text{ring}(\varphi,\theta) -i\sin\gamma~\textbf{K}_\text{prec}(\varphi,\theta)\bigg) \:.
\label{eq_EFF}
\end{align}
We notice that the first term on the right-hand side resembles the vector diffraction of a 3D-ring multiplied by $\cos\gamma$, cf. Eq.~(\ref{equ:nxE_ring}). The additional second term on the right-hand side exists only for $\gamma > 0$ (conical cavities) and arises from the precession of the electric field along its trajectory around the cone axis. This term is phase shifted by $\pi/2$ w.r.t. the first term as indicated by the prefactor $i$. 
The above-mentioned reasoning resembles, in a number of aspects, the action of geometric phases known in various examples throughout mesoscopic physics of electrons and photons and beyond~\cite{geomphaseBook_1989}.
In particular we find the cone's opening angle $2\gamma$ to play an important role, as expected for example from electronic transport in inhomogeneous magnetic fields~\cite{Loss_1990,Nagasawa_2013,Frustaglia2001,Hentschel2004}. In idealized one-dimensional electronic magnetotransport the geometric phase is known to be directly related to the cone's opening angle (more precisely, it is half (for electronic spin) of the solid angle spanned by the spin dynamics in parameter (magnetic-field) space as the ring trajectory is traversed). However, Eq.~(\ref{eq_EFF}) proofs things to be more complex in the generic (and three-dimensional) situation considered here where in particular an additional transition to the far field has to be taken into account\\
In the same fashion as in section~\ref{sec:diff_ring}, we derive the far-field components $E_\varphi$ and $E_\theta$:
\begin{align}
E_\varphi &={} -\textbf{K}\cdot\textbf{e}_\theta =	 \cos\gamma\left(\left(K_\text{x,ring}\cos\varphi + K_\text{y,ring}\sin\varphi \right)\sin\theta - K_\text{z,ring}\cos\theta\right) \nonumber\\
				& \quad\quad +i\sin\gamma \left(\left(K_\text{x,prec}\cos\varphi + K_\text{y,prec}\sin\varphi \right)\sin\theta - K_\text{z,prec}\cos\theta\right) \\
				&={} \cos\gamma~E_{\varphi,\text{ring}} + i\sin\gamma~E_{\varphi,\text{prec}} \\[10pt]
E_\theta &={} \textbf{K}\cdot\textbf{e}_\varphi =   \cos\gamma\left(-K_\text{x,ring}\sin\varphi + K_\text{y,ring}\cos\varphi\right) \nonumber\\ 
				& \quad\quad + i\sin\gamma\left(-K_\text{x,prec}\sin\varphi + K_\text{y,prec}\cos\varphi\right) \\
				&={} \cos\gamma~E_{\theta,\text{ring}} + i\sin\gamma~E_{\theta,\text{prec}} \:,
\end{align}
and notice that both components $E_\varphi$ and $E_\theta$ undergo a phase shift that arises from the precession of the electric field vector inside the conical cavity. This is the very origin of the phase 
$\delta$ between the far-field components $E_\varphi$ and $E_\theta$ which in turn results in a change of the orientation angle $\Psi$, cf.~Eq.~(9) of the manuscript.\\
For the sake of completeness, we examine the components of the diffraction integral in Eq.~(\ref{eq_EFF}). To this end, we express the electric field vector of the TE-like WGM by the unit vector parallel to the cone surface and by the azimuthal unit vector: $\textbf{E}(u',\phi')=\left(\tilde{A}_\parallel(u')\textbf{e}_\parallel(\phi') + \tilde{A}_\phi(u')\textbf{e}_\phi(\phi') \right)\exp{\left(i m\phi'\right)}$ with $\textbf{e}_\parallel = -\sin\gamma~\textbf{e}_\text{r}+\cos\gamma~\textbf{e}_\text{z}$, cf. Fig.~\ref{fig:S1}. Thus, the components of $\textbf{K}_\text{ring}$ and $\textbf{K}_\text{prec}$ can be written as:
\begin{align}
K_\text{x,ring} &={}  \iint\cos\gamma~\tilde{A}_{\parallel}(u')\sin\phi'\exp{(im\phi')}\exp{\left(-i\textbf{k}\cdot\textbf{x}'\right)} \d a' \\
K_\text{y,ring} &={}  -\iint\cos\gamma~\tilde{A}_{\parallel}(u')\cos\phi'\exp{(im\phi')}\exp{\left(-i\textbf{k}\cdot\textbf{x}'\right)} \d a' \\
K_\text{z,ring} &={}  \iint~\tilde{A}_{\phi}(u')\exp{(im\phi')}\exp{\left(-i\textbf{k}\cdot\textbf{x}'\right)} \d a' \\
K_\text{x,prec} &={}  \iint\tilde{m}\left(\sin\gamma~\tilde{A}_{\parallel}(u')\cos\phi' + \tilde{A}_\phi(u')\sin\phi'\right)\exp{(im\phi')}\exp{\left(-i\textbf{k}\cdot\textbf{x}'\right)} \d a' \\
K_\text{y,prec} &={}  \iint\tilde{m}\left(\sin\gamma~\tilde{A}_{\parallel}(u')\sin\phi' - \tilde{A}_\phi(u')\cos\phi'\right)\exp{(im\phi')}\exp{\left(-i\textbf{k}\cdot\textbf{x}'\right)} \d a' \\
K_\text{z,prec} &={}  -\iint\tilde{m}\cos\gamma~\tilde{A}_{\parallel}(u')\exp{(im\phi')}\exp{\left(-i\textbf{k}\cdot\textbf{x}'\right)} \d a'
\end{align}
where $\tilde{m}=m-\textbf{k}\cdot\frac{\partial\textbf{x}'}{\partial \phi'}= m - k(R-u'h\sin\gamma)\sin(\phi'-\varphi)\cos\theta$.\\
The parametrization $\textbf{x}'$ of the surface $S$ representing the inclined cone surface reads
\begin{align}
\textbf{x}'(\phi',u') &= R \textbf{e}_\text{r}(\phi') + u' h \textbf{e}_\parallel(\phi') =  R \textbf{e}_\text{r}(\phi') + u' h\cos\gamma~ \textbf{e}_\text{z} - u'h\sin\gamma~\textbf{e}_\text{r} \\
&= \left(R-u'h\sin\gamma \right)\textbf{e}_\text{r} + u'h\cos\gamma~\textbf{e}_\text{z} \:.
\end{align}
Evaluating the term $\textbf{k}\cdot\textbf{x}'$ yields thus
\begin{align}
\textbf{k}\cdot\textbf{x}' &= k\left(R-u'h\sin\gamma\right)\left(\cos\theta\cos\varphi\cos\phi' + \cos\theta\sin\varphi\sin\phi'  \right) + u'h k\cos\gamma\sin\theta\\
&= \tilde{k}_1\cos\left(\phi'-\varphi\right) +\cos\gamma~\tilde{k}_2u' - \sin\gamma~kh\cos\theta\cos\left(\phi'-\varphi\right)u' \:,
\end{align}
where the additional third term on the right-hand side arises from the inclination of the wall, cf. Eq.~(\ref{equ:kx_ring}), and depends on $\phi'$ and $u'$. As a result, the components of the diffraction formula do not factorize into an $\phi'$- and $u'$-integration as in the case of the 3d-ring. Note that the $u'$-integrations can not be treated approximately as a mere Fourier-Transform of the amplitudes as in the case of the 3d-ring. We demonstrate this exemplarily for $K_\text{x,ring}$:
\begin{align}
K_\text{x,ring} &={} h R  \int_{-1/2}^{1/2}  \cos\gamma~\tilde{A}_{\parallel}(u')F_\text{x}(u')\exp{\left(-i\cos\gamma~\tilde{k}_2u'\right)}  \d u' \\
F_\text{x}(u') &={} s(u') \int_{0}^{2\pi}\sin\phi'\exp{\left(i m \phi'\right)}\exp{\left(-i\left(\tilde{k}_1-\sin\gamma~kh\cos\theta~u'\right)\cos\left(\phi'-\varphi\right)  \right)\d \phi'} \:,
\end{align} 
where $s(u')=\left(1-u'\frac{h}{R}\sin\gamma\right)$. We discover a more complicated case: the amplitude is modulated by the function $F_\text{x}(u')$ already causing a different phase relation of the far-field components. In addition to the above mentioned phase change due to precession, the far-field polarization quantities $\Psi$ and $\chi$ depend also on the amplitude profiles $\tilde{A}_\parallel(u')$ and $\tilde{A}_\phi(u')$.\\
%



\bibliography{references_SM.bib}